\documentclass[9pt,superscriptaddress,amsmath,amssymb,aps,twocolumn]{revtex4-2}
\usepackage{graphicx}
\usepackage{breakurl}
\usepackage[a4paper, top=2cm, bottom=2cm, left=1.5cm, right=1.5cm]{geometry}
\usepackage[colorlinks=true,linkcolor=blue,urlcolor=blue,citecolor=blue,anchorcolor=blue]{hyperref}

\begin{document}

\title{Sideward contact tracing and the control of epidemics in large gatherings}

\author{Marco Mancastroppa}
\affiliation{Dipartimento di Scienze Matematiche, Fisiche e Informatiche,
 Universit\`a degli Studi di Parma, Parco Area delle Scienze, 7/A 43124 Parma, Italy}
\affiliation{INFN, Sezione di Milano Bicocca, Gruppo Collegato di Parma, Parco Area delle Scienze, 7/A 43124 Parma, Italy}
\author{Andrea Guizzo}
\affiliation{Dipartimento di Scienze Matematiche, Fisiche e Informatiche,
 Universit\`a degli Studi di Parma, Parco Area delle Scienze, 7/A 43124 Parma, Italy}
\affiliation{INFN, Sezione di Milano Bicocca, Gruppo Collegato di Parma, Parco Area delle Scienze, 7/A 43124 Parma, Italy}
\author{Claudio Castellano} 
\affiliation{Istituto dei Sistemi Complessi (ISC-CNR), Via dei Taurini 19, I-00185 Roma, Italy}
\author{Alessandro Vezzani}
\affiliation{Istituto dei Materiali per l'Elettronica ed il Magnetismo (IMEM-CNR), Parco Area delle Scienze, 37/A 43124 Parma, Italy}
\affiliation{Dipartimento di Scienze Matematiche, Fisiche e Informatiche,
 Universit\`a degli Studi di Parma, Parco Area delle Scienze, 7/A 43124 Parma, Italy}
 \affiliation{INFN, Sezione di Milano Bicocca, Gruppo Collegato di Parma, Parco Area delle Scienze, 7/A 43124 Parma, Italy}
\author{Raffaella Burioni}
\affiliation{Dipartimento di Scienze Matematiche, Fisiche e Informatiche,
 Universit\`a degli Studi di Parma, Parco Area delle Scienze, 7/A 43124 Parma, Italy}
\affiliation{INFN, Sezione di Milano Bicocca, Gruppo Collegato di Parma, Parco Area delle Scienze, 7/A 43124 Parma, Italy}


\begin{abstract}
Effective contact tracing is crucial to contain epidemic
spreading without disrupting societal activities especially in the
present time of coexistence with a pandemic outbreak. Large gatherings
play a key role, potentially favouring superspreading events.
However, the effects of tracing
in large groups have not been fully assessed so far. We show that beside forward tracing, which reconstructs to whom disease spreads, 
and backward tracing, which searches from whom disease spreads, 
a third “sideward" tracing is always present, when tracing gatherings. 
This is an indirect tracing that detects infected asymptomatic
individuals, even if they have neither
been directly infected by, nor they have directly transmitted the
infection to the index case. We analyse this effect in a model of
epidemic spreading for SARS-CoV-2, within the framework of simplicial
activity-driven temporal networks. We determine the contribution
of the three tracing mechanisms to the suppression of epidemic
spreading, showing that sideward tracing induces a non-monotonic behaviour 
in the tracing efficiency, as a function of the size of the gatherings. Based on our results,
we suggest an optimal choice for the sizes of the gatherings to be traced and we test the
strategy on an empirical dataset of gatherings in a University Campus.
\end{abstract}

\maketitle

\section{Introduction}
Public debate about measures to curb pandemic spreading in the present
time, where new extensive outbreaks of SARS-CoV-2 are emerging due to highly transmissible variants, is
dominated by opposite tensions toward contrasting goals. On the one
hand many advocate the continued implementation of strict policies
aimed at preventing virus propagation and all the resulting
consequences in terms of deaths, sufferings and long-lasting health
problems~\cite{Moghadas2021,didomenico2021control,Scott2020}. On the other hand the heavy
economic, societal and psychological costs of restrictions push many
to call for lifted restrictions and a general rapid
return to normal activities~\cite{Bonaccorsi2020,Mandel2020,McKee2020}. Finding an acceptable tradeoff between these two clashing tendencies is
extremely hard. Decisions must necessarily be taken by
policy-makers, but it is the task of science to provide informed
advice and predict likely scenarios~\cite{chang2021mobility,didomenico2021control,Moghadas2020,Pullano2021}.

A key role in epidemic spreading is
played by large gatherings of people, which are fundamental
structures of social networks~\cite{sekara2016fundamental,battiston2020networks,cencetti2021temporal}.
Indeed one of the most common measures taken by governments to limit
disease propagation is the prohibition to hold gatherings
above a given size, to suppress the chance
of 'superspreading events' (SSEs), where a large number of people are
simultaneously infected~\cite{lewis2021superspreading,althouse2021superspreading,adam2020clustering,majira2021super}.
A firm theoretical underpinning for these types of decisions is still lacking.
Recent results have shown that large gatherings may have dramatic effects when 'complex contagion' is at work,
i.e. the probability of transmission depends nonlinearly on the
number of independent exposures an individual is subjected to~\cite{iacopini2019simplicial}.
This is however not the typical case for disease spreading (simple
contagion), where contact with two infectious individuals
simply doubles the contagion probability.
Nevertheless the meeting of large groups implies the formation of a
quadratically large number of contacts and for this reason limitations
to the size of allowed gatherings has a substantial impact on
epidemic propagation~\cite{stonge2021social}.
In this paper, we show that the specific nature that makes large
gatherings threatening, on the other hand can
improve the effectiveness of contact tracing procedures
and therefore control the spreading process without disrupting societal
activities.

The tracing of contacts of infected individuals and their subsequent
isolation is one of the main weapons against disease spreading,
in particular when pre-symptomatic and asymptomatic individuals are responsible for a large share of transmission events~\cite{Fraser2004,Thompson2016}. In the absence of contact tracing, all presymptomatic and asymptomatic infections can go undetected as observed with SARS-CoV-2~\cite{Pinotti2020,Pullano2021,Moghadas2020,michael2021}.
In the usual framework where only binary contacts occur, 
two types of contact tracing (CT) are possible when somebody
is found infected:
backward CT follows the transmission chain back in time, aiming at
reconstructing who infected the index case~\cite{kojaku2021effectiveness,Bradshaw2021,endo2020implication};
forward CT instead looks for people potentially infected upon
contact with the index case prior to his/her detection~\cite{kojaku2021effectiveness,Bradshaw2021,endo2020implication}.
In the case of a gathering, if the group is traced as a whole, the efficacy of the
procedure is greatly enhanced because of an indirect effect that
we dub \textit{sideward CT}, qualitatively different from the two
other tracing types.
Sideward CT allows to indirectly trace infected asymptomatic individuals that have participated
in the same gathering, even if they have neither been directly
infected by the index case nor they have directly infected him/her.
We analyze this effect by considering a model of epidemic spreading
(tailored to describe SARS-CoV-2 transmission) on a temporal network with groups activation~\cite{petri2018simplicial}, a standard
modelling scheme for mutually interacting groups of people evolving with time.
In this framework we determine analytically the contribution of
each of the three types of tracing to the suppression of epidemic spreading,
revealing the great importance of sideward tracing. We show that due to the simultaneous 
contribution of the three mechanisms, the effectiveness of tracing gatherings is non-monotonic 
as a function of the gathering size. In particular, the effectiveness features a maximum, where
the contact tracing including all the three mechanism is maximally effective.
We test our results on synthetic group size distributions and on
an empirical dataset for gatherings in a University Campus, whose group
distribution is determined via WIFI data, and we suggest an optimal strategy to choose
the typical size of the gatherings to be traced.
This optimal tracing of large gatherings can readily offset
their potential for disease spreading, thus providing a concrete strategy
to curb epidemics still allowing activities to proceed with limited restrictions.

\section{Results}

\subsection{Epidemic spreading on simplicial temporal networks}

\begin{figure}
\centering
\includegraphics[width=88mm]{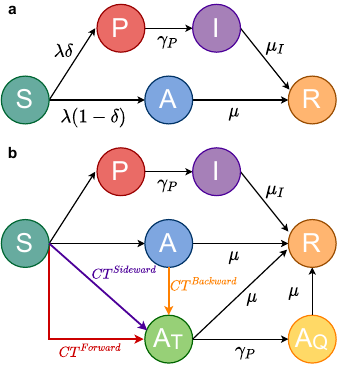}\\
 \caption{{\bf The epidemic compartmental model with and without
 tracing.} {\bf a} Scheme of the transitions in the compartmental epidemic model
 without CT. {\bf b} The enlarged compartmental model in the
 presence of forward, backward and sideward CT. The transition
 rates for infections and for CT are discussed in details in the
 text and in the Supplementary Material, SM.}
 \label{fig:figure1}
\end{figure}

Temporal networks constitute a general framework developed in recent years, to model systems in which interactions among elements are continuously added and removed, typically on the same temporal scale of the process that is considered on the network~\cite{holme2019}. 
In particular, in activity-driven networks the activation of the links is determined by the activity of the nodes~\cite{Perra2012,Ubaldi2016,Mancastroppa2019}. 
Here we focus on activity-driven simplicial temporal networks, in which the temporal dynamics of social contacts occurs in groups, accounting for the higher-order nature of interactions~\cite{petri2018simplicial,battiston2020networks,cencetti2021temporal}.

In particular, we consider individuals as nodes of a temporal network, 
that interact with others by taking part in gatherings, modelled as
{\it simplices}: each node in a gathering is in contact with
all the others forming thus a fully connected cluster.
The interaction network evolves by activating
simplices at rate $a$, the simplex activity;
when a $(s-1)$-simplex (i.e. a clique of size $s$) is active, $s$ nodes
are chosen uniformly at random to participate in the simplex, producing
$s(s-1)/2$ interactions. 
Then the links are destroyed and the process is iterated.
The simplex size $s$ is drawn from a distribution $\Psi(s)$
that models the heterogeneity in the size of gatherings.
Usual pairwise interactions correspond to simplices with $s=2$.
On top of this network, the pathogen spreading dynamics
is described by a compartmental model (see
Fig.~\ref{fig:figure1}\textbf{a}) accounting for the main stages
of a contagious disease with asymptomatic and pre-symptomatic
transmission, such as SARS-CoV-2
infection~\cite{michael2021,Guan2020,who_china_report}.
A susceptible individual $S$ is infected with probability $\lambda$
if connected with an infected node by a link of the temporal simplex.
Infected individuals branch out in two paths: the infection can be symptomatic
with probability $\delta$, $S \xrightarrow[]{\lambda \delta} P$, or
asymptomatic with probability
$(1-\delta)$, $S \xrightarrow[]{\lambda (1-\delta)} A$.
Asymptomatic nodes $A$ spontaneously recover ($R$)
with rate $\mu=1/\tau$; pre-symptomatic individuals
spontaneously develop symptoms with rate $\gamma_P=1/\tau_P$, thus
with a Poissonian process $P \xrightarrow[]{\gamma_P} I$, and then
infected symptomatic $I$ spontaneously recover with rate
$\mu_I=\mu \gamma_P/(\gamma_P -\mu)$. The average infectious period is therefore
$\tau$ both for asymptomatic and symptomatic individuals. Susceptible
$S$, recovered $R$, asymptomatic $A$ and pre-symptomatic $P$
individuals have an equal uniform probability to join a simplex.
Over this temporal network we introduce an adaptive behavior, assuming that
symptomatic individuals $I$ are immediately isolated and thus are not
able to participate in simplices and propagate the infection~\cite{Vankerkove2013,Pozzana2017,Mancastroppa2020,mancastroppa2021contacttracing}.

SIR compartmental models are characterised by an active phase, in which the epidemic propagates to a finite fraction of nodes, and an absorbing phase with the number of infected individuals exponentially decaying to zero. The epidemic threshold separating the two phases can be analytically obtained by means of a linearisation procedure, providing the stability of the absorbing phase. In particular in activity-driven models the calculation of the threshold is exact since local correlations are continuously destroyed by link reshuffling
\cite{Perra2012,Ubaldi2016,Mancastroppa2019}. 
We use the infection probability $\lambda$ as control parameter; i.e. only above its critical value $\lambda_C$ -- the epidemic threshold -- extensive outbreaks
reach a finite fraction of the population. The effectiveness of a control
strategy can be estimated by the increase of $\lambda_C$. In
the simplicial temporal network with no adaptive measures (NA)
the threshold is~\cite{petri2018simplicial}:
\begin{equation}
\lambda_C^{NA}=\frac{\mu}{a \langle s(s-1) \rangle},
\label{eq:thresh_NA}
\end{equation}
where $\langle f(s) \rangle$ denotes the average over the different simplex sizes $\langle f(s) \rangle = \int ds \, \Psi(s) f(s)$. Notice that the critical condition can be reformulated in terms of the basic reproduction number $R_0$, which is defined as $R_0=\lambda/\lambda_C^{NA}=\lambda a \langle s(s-1) \rangle/\mu$. 

If symptomatic individuals $I$ are immediately and perfectly isolated once they develop symptoms, the epidemic threshold is (see the Supplementary Material - SM, Section I.D for details):
\begin{equation}
\lambda_C^{sympto}=\lambda_C^{NA} \frac{\frac{\gamma_P}{\mu}}{\delta+\frac{\gamma_P}{\mu}(1-\delta)}.
\label{eq:thresh_isolation}
\end{equation}
Also in this case the concept of epidemic threshold can be reframed in terms of the reproduction number $R_0$, which is defined as $R_0=\lambda/\lambda_C^{sympto}=\frac{\lambda a \langle s(s-1) \rangle}{\mu} \frac{\delta+(1-\delta) \gamma_P/\mu}{\gamma_P/\mu}$.  In Eq. \eqref{eq:thresh_isolation} for $\delta=0$ (no symptomatic infection) we recover Eq. \eqref{eq:thresh_NA}, while for $\delta=1$ the epidemic transmission is maximally reduced due to isolation of symptomatic individuals. This increases $\lambda_C$ of a factor $\frac{\gamma_P}{\mu}$.
Hereafter, $\lambda_C^{sympto}$ will be the reference for the evaluation of the performance of CT strategies.

\subsection{Contact tracing mechanisms in simplices}
We consider a traditional (non app-based) CT
process~\cite{mancastroppa2021contacttracing} on simplices:
the goal is to identify and isolate infected asymptomatic individuals.
Note that since individuals in state $I$ are instantaneously isolated,
they do not participate in simplices. 
Only individuals in state $P$ or $A$ may spread the infection. Within this assumption presymptomatic individuals also represent people 
with no symptoms or very weak ones, thus still able to attend events,
at least for a certain time before isolation (imperfect isolation).
CT is activated when a presymptomatic individual develops significant
symptoms $P \to I$: in such a case each simplex
he/she has participated in during the
previous $T_{CT}$ days is traced as a whole, with a probability $\epsilon(s)$.
Each node belonging to a traced simplex is tested and, if found in
the asymptomatic infected (A) state, isolated. The tracing procedure is stopped at the first-step, so we do not consider iterative tracing. The probability of tracing a simplex
can depend in general on the size $s$:
people typically remember only some of the
gatherings they joined; some simplices are easily
fully traced (e.g. school classes, workplace meetings) while
others are not easily reconstructed (e.g. interactions on public
transportation, shops, restaurants).
The dependence of $\epsilon(s)$ on the simplex size $s$ also allows
to model tracing strategies targeted at groups of a given size. 

In the present framework CT is modelled by introducing two additional
compartments for asymptomatic individuals (see Fig.~\ref{fig:figure1}\textbf{b}):
traced asymptomatic, $A_T$, and quarantined asymptomatic, $A_Q$.
An $A_T$ node is asymptomatic and infective, just as an $A$ individual,
but it has been in touch with a pre-symptomatic individual who remembers
the gathering where the contact took place. 
We assume that tracing time is shorter than the typical time of the epidemic evolution, so that it can be considered instantaneous.
Therefore, when the pre-symptomatic develops symptoms,
the $A_T$ node enters quarantine and becomes an $A_Q$ node,
which is isolated and hence does not participate in simplices.
We describe the effect of isolation of contacts occurring when the index node develops symptoms setting the transition $A_T \to A_Q$ at the same rate
$\gamma_P$ for the appearance of symptoms (see \cite{mancastroppa2021contacttracing} for a discussion on the effects of longer tracing times).

Besides these assumptions, our model involves other hypotheses in the tracing procedure and in the epidemic transmission. For example, we assume that all nodes within a traced simplex are identified and that, once the nodes are found infected, the isolation of quarantined individuals is perfect. We also assume that symptomatic and asymptomatic individuals have the same infectivity and that the probability of infection is independent of the simplex size. All these assumptions can be relaxed in the model in a straightforward manner, by including new parameters or interdependences between the parameters with no conceptual obstacles. We do not expect this to modify substantially our findings on the general behaviour of the tracing procedure that we are going to discuss.

\begin{figure*}
\centering
\includegraphics[width=180mm]{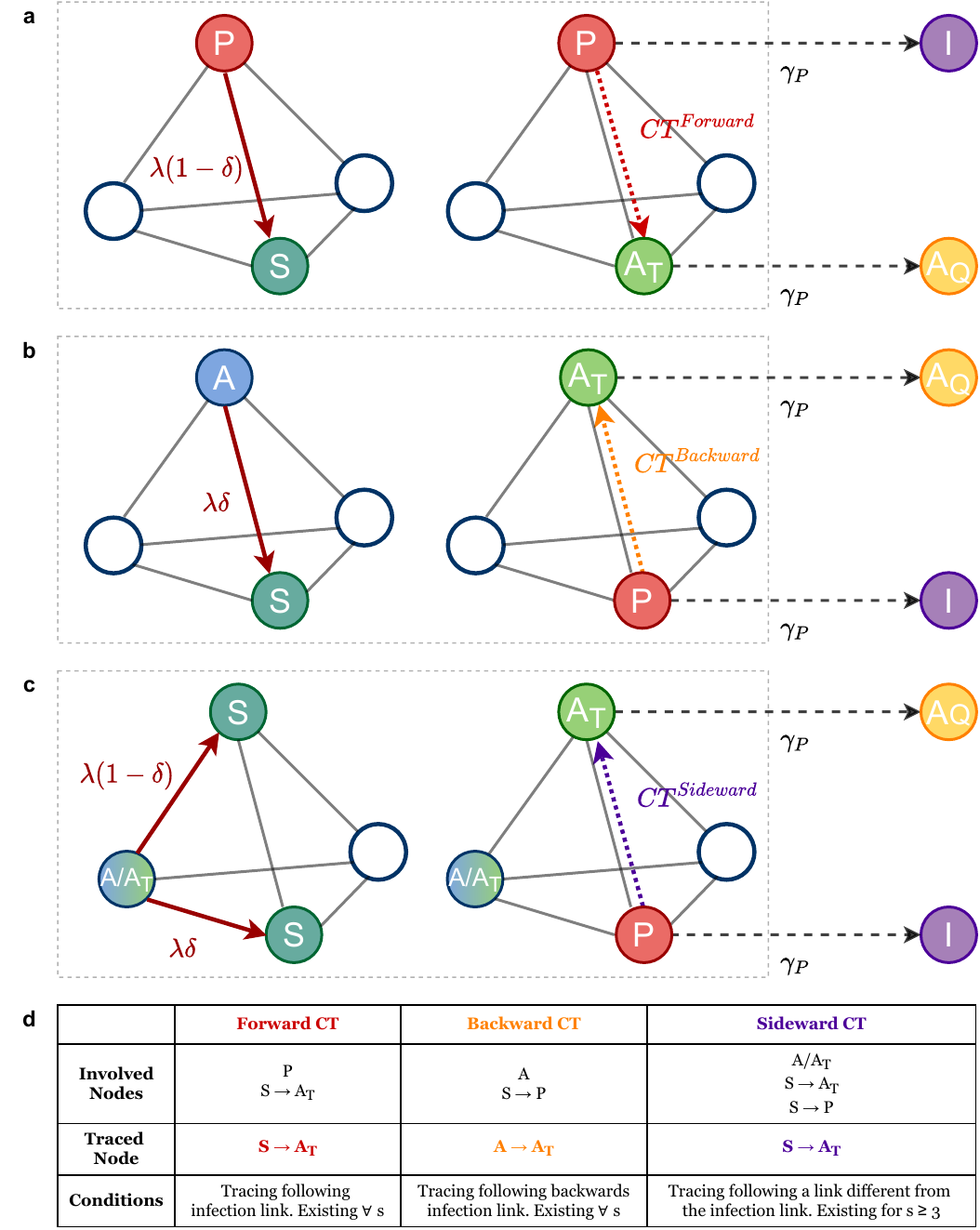}\\
 \caption{{\bf Contact tracing mechanisms in simplices: forward,
 backward and sideward CT.} {\bf a} Scheme of the forward
 CT. {\bf b} Scheme of the backward CT. {\bf c} Scheme of the
 sideward CT: note that in this case the CT can be activated both
 if the infected node is $A$ or $A_T$. Moreover in this simplex
 also backward CT can occur on the node $A$, if present, activated
 by $S \to P$, but for simplicity we do not show it in the
 scheme. Finally, sideward CT can be activated if the asymptomatic
 and symptomatic infections are produced by two different
 asymptomatic individuals (or traced asymptomatics), who
 participate in the simplex; this process is considered in the
 model (see SM) and produce a term quadratic in the density of
 asymptomatic nodes. However for simplicity in the scheme we
 consider the first-order term, which is the only one to survive
 the linearisation for the calculation of the epidemic
 threshold. In all the tracing mechanisms, the traced node becomes $A_T$ upon contact with the index case and then is isolated $A_T \to A_Q$ at the same rate $\gamma_P$ for the appearance of symptoms in the index case $P \to I$, assuming the tracing time shorter then the typical time of the epidemic evolution, so that it can be considered instantaneous. {\bf d}  Summary of the main features of the three CT
 mechanisms.}
 \label{fig:figure2}
\end{figure*}

Inside simplices three types of contact tracing mechanisms are at work:
forward and backward CT, already present for pairwise
interactions~\cite{mancastroppa2021contacttracing,endo2020implication,kojaku2021effectiveness,Bradshaw2021},
are augmented by
sideward CT, which is specific of simplices traced as a whole.

\textit{Forward CT} looks for asymptomatic individuals infected by a
presymptomatic index case who, upon developing symptoms, activates the CT.
This mechanism involves, in the simplex, one susceptible node and a presymptomatic
index case (see Fig.~\ref{fig:figure2}\textbf{a}).
At the moment of the infection event, the susceptible node becomes asymptomatic and traced ($A_T$),
the tracing occurs along the same link on which the infection takes place.
See Methods for the associated term in the mean-field equations.
 
\textit{Backward CT} is the search for the source of infection of a
symptomatic index case, who activates the CT. This mechanism is at
work when in a simplex an asymptomatic node infects a
susceptible node making it presymptomatic (see Fig.~\ref{fig:figure2}\textbf{b}):
the asymptomatic node is traced (becoming $A_T$) along the
contact that produced the infection, but the tracing goes in the
opposite direction. The asymptomatic node
traced in the process is already infected when he/she enters the
simplex, thus the tracing occurs after the infection event and after he/she
has potentially infected other nodes. See Methods for the associated term in the mean-field equations.

\textit{Sideward CT} finds asymptomatic nodes infected by {\em other}
asymptomatic individuals, by exploiting the presence in the simplex of
a third node that develops symptoms (see Fig.~\ref{fig:figure2} \textbf{c}).
Therefore, sideward CT occurs when in the simplex there is at least
one asymptomatic $A$ (or traced asymptomatic $A_T$) node, that infects
a susceptible node (that becomes asymptomatic) and also infects at
least another susceptible node (that becomes presymptomatic). The
participation in the same simplex implies that the asymptomatic is
traced so that it enters quarantine when the presymptomatic develops
symptoms and activates CT. Another possibility is that the two
susceptible nodes in Fig.~\ref{fig:figure2} \textbf{c} are infected by different
asymptomatic individuals, however the term associated to this process
is quadratic in the density of asymptomatic nodes and hence does not
influence the threshold value. Sideward CT
requires the presence of at least three nodes, thus it is active only
if $s \geq 3$; moreover it occurs "laterally", since the traced
contact does not transmit the infection. Note that in the same simplex
also backward CT can be activated on the infector if it is an
infected asymptomatic $A$. In the linearised mean-field equations (see
Methods), sideward CT is described by a transition from $S$ directly
to $A_T$ with the following term:
\begin{widetext}
\begin{equation}
C^{Sideward}=a \lambda (1-\delta) \left\langle \epsilon(s) s (s-1) \left[ 1- (1-\lambda \delta)^{(s-2)}\right] \right\rangle [A(t)+A_T(t)],
\label{eq:CTside}
\end{equation}
\end{widetext}
where $a$ is the simplex activation rate, the probability to take part
to the simplex for the susceptible node is proportional to $s$ and, in the linearised regime, the
probability that among the remaining $(s-1)$ nodes there is an
asymptomatic (or a traced symptomatic) node is $(s-1) [A(t)+A_T(t)]$.
The asymptomatic infection occurs with probability
$\lambda(1-\delta)$ and $\left[ 1- (1-\lambda \delta)^{(s-2)}\right]$
is the probability that, at least in one of the remaining $(s-2)$ nodes,
a symptomatic contagion occurs. $\epsilon(s)$ is the probability
that the simplex is traced and the term is averaged over all simplex sizes.

The CT terms feature a highly non-trivial dependence on the control parameter $\lambda$ (see Eq.~\eqref{eq:CTside}, Methods and SM): this complicates the calculation of the epidemic threshold $\lambda_C$ (see SM) and prohibits to formulate the reproduction number $R_0$ as a simple function of $\lambda$, unlike the case with only symptomatic isolation. Thus, hereafter we will describe the critical behaviour by means of the epidemic threshold $\lambda_C$. Again, the goal of CT is to increase the threshold: the more effective the CT strategy, the higher the value of $\lambda_C$.

\subsection{The effects of forward, backward and sideward tracing in simplices}

\begin{figure*}
\centering
\includegraphics[width=180mm]{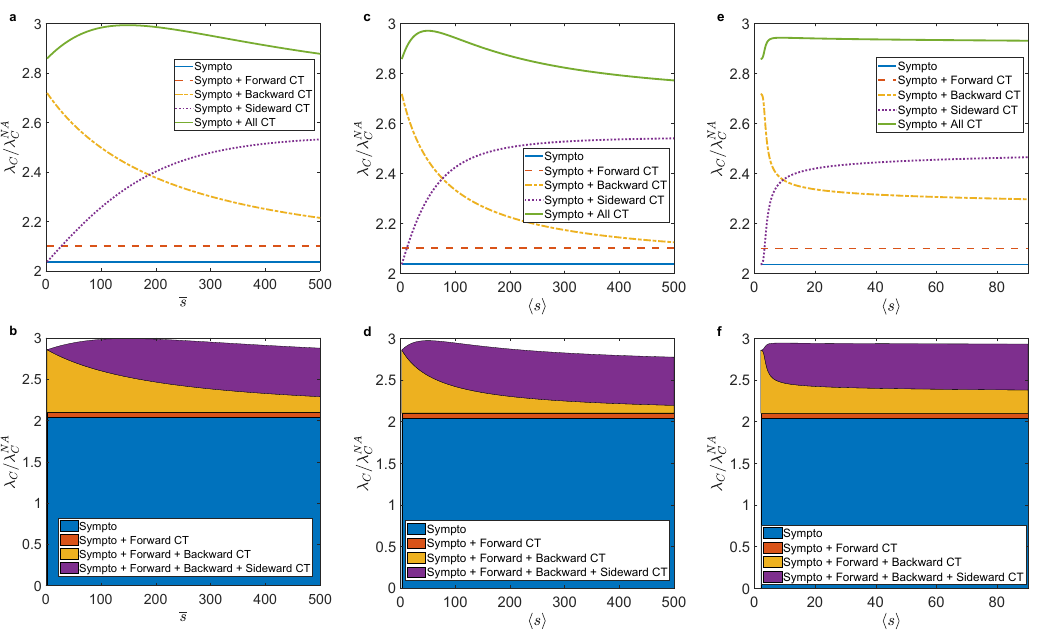}
 \caption{{\bf The effects of forward, backward and sideward contact
 tracing.} {\bf a} We consider $\Psi(s)=\delta(s-\overline{s})$
 and we plot, as a function of $\overline{s}$, the ratio between
 the epidemic threshold $\lambda_C$ when symptomatic isolation and
 CT are activated and the epidemic threshold $\lambda_C^{NA}$ of
 the non-adaptive case. We plot curves corresponding to the activation of
 each CT mechanism alone and then to the case when they are all active. {\bf b}
 For the same $\Psi(s)$ of panel {\bf a} we plot the ratio
 $\lambda_C/\lambda_C^{NA}$ as a function of $\overline{s}$ for
 symptomatic isolation and activating progressively all CT
 mechanisms.
 {\bf c} and {\bf d}:
 As in panels {\bf a} and {\bf b}, respectively, but for
 $\Psi(s) \sim e^{-\beta s}$, with $s \in[2,\infty)$.
 {\bf e} and {\bf f}: 
 As in panels {\bf a} and {\bf b}, respectively, but for
 $\Psi(s) \sim s^{-(\nu+1)}$ and $s \in [2,500]$.
 In panels {\bf c-d} and {\bf e-f} we change the average simplex
 size $\langle s \rangle$, by varying $\beta$ and $\nu$,
 respectively, thus changing the tails of the $\Psi(s)$ distribution.
 In all panels $\epsilon(s)=0.3 \, \forall s$.
 All the other parameters are fixed as discussed in the
 Methods.}
 \label{fig:figure3}
\end{figure*}

The three tracing mechanisms contribute
differently to the epidemic mitigation and their effectiveness depends
on the structure of interactions, i.e. the simplex distribution $\Psi(s)$.
To compare their performance we assume that the probability
to be traced is equal for simplices of any size,
i.e. $\epsilon(s)=\epsilon, \, \forall s$.
We keep constant $\overline{n}=a\langle s(s-1) \rangle$, the
average number of contacts per individual per time unit,
so that different distributions $\Psi(s)$ correspond to the same number
of interactions, arranged in simplices of different size.
We assess the specific effect of each CT
by calculating the threshold
first when only symptomatic individuals are isolated, then with each CT
mechanism at work separately and finally when all CTs are active (see Methods and SM).
In Fig.~\ref{fig:figure3} \textbf{a}-\textbf{b} we
consider all simplices of the same size $\overline{s}$, i.e.
$\Psi(s)=\delta(s-\overline{s})$, where $\delta (x)$ is the Dirac delta-function, while in Fig.~\ref{fig:figure3} \textbf{c}-\textbf{d}
an exponential distribution $\Psi(s) \sim e^{-\beta s}$ is considered, with
$s \in [2,\infty)$.
Finally, in Fig.~\ref{fig:figure3} \textbf{e}-\textbf{f} the simplex size distribution is a
power-law $\Psi(s) \sim s^{-(\nu+1)}$, with $s \in [2,s_M]$
and $s_M$ being of the order of hundreds of nodes, as often observed in real
systems~\cite{sekara2016fundamental}.
The effects of the isolation of symptomatics and of forward CT are
independent of $\Psi(s)$, while backward and sideward CT strongly depend on
$\Psi(s)$. Sideward tracing is very effective
in the presence of large simplices, where lateral tracing is more
probable,  while it cannot be triggered for $s=2$
(i.e. simple links). In large
simplices, sideward CT is indeed able to trace all new asymptomatic
individuals and isolate them at the infection,
avoiding epidemic spreading and the explosive effects of SSEs.
On the other hand, backward tracing is more effective on small simplices:
in large clusters, with many contacts, it only traces the source of
infection, while other simultaneous contagions may occur and
go undetected.

Interestingly, when all CT mechanisms are at work, the combination of
backward and sideward tracing gives rise to a non-monotonic behaviour
as a function of the simplex size. In particular, the
threshold features a maximum, where CT is maximally effective.
In networks with a single simplex size or with a sharp exponential
distribution, the size corresponding to this maximum is of the
order of $100$ nodes, while for broader distributions, with very large
clusters dominating the transmission even at small $\langle s \rangle$,
the tracing is maximally effective when $\langle s \rangle \approx 10$.
The position of this maximum strongly depends also on the fraction of
asymptomatic nodes (see SM). In particular, a large fraction of asymptomatic nodes is most effectively traced by sideward tracing while for few asymptomatics backward tracing is most effective.

\subsection{Optimal strategies for contact tracing in a real setting}
\begin{figure*}
\centering
\includegraphics[width=180mm]{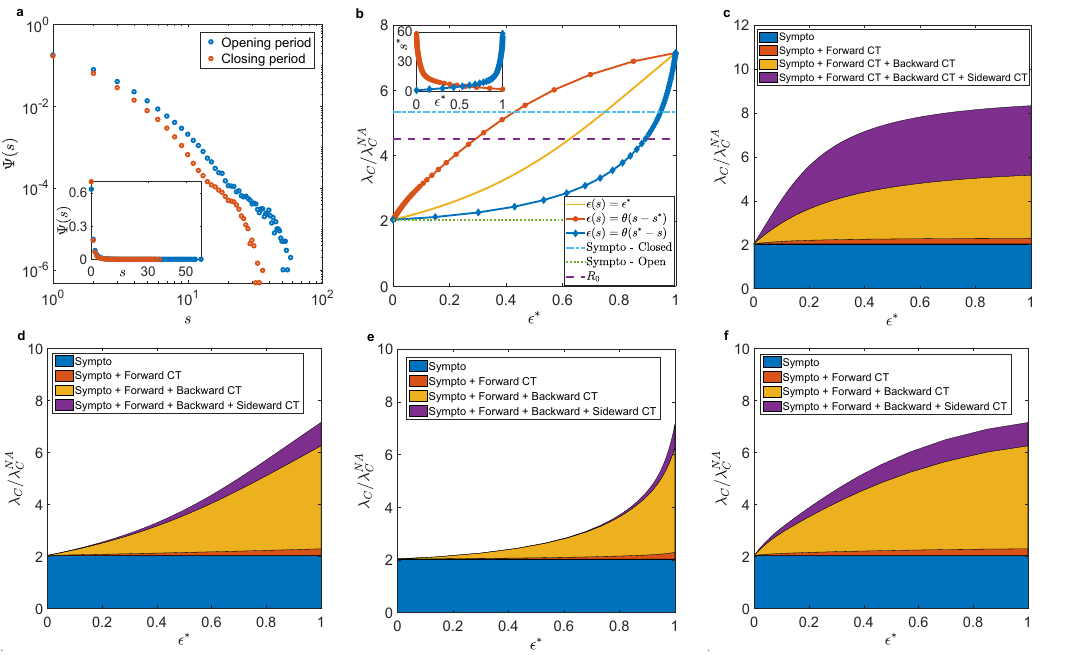}
 \caption{{\bf Tracing strategies on empirical University gatherings
 data.} {\bf a} We plot the distributions $\Psi(s)$ obtained via
 WIFI data for the University of Parma during the partial opening and
 closure periods. The main plot is in log-log scale, while in the
 inset the same distributions are plotted in linear scale.
 {\bf b} We plot the ratio
 $\lambda_C/\lambda_C^{NA}$ as a function of
 $\epsilon^*=\frac{\langle \epsilon(s) (s-1)\rangle}{\langle s-1 \rangle}$, where $\lambda_C^{NA}$ is the epidemic threshold in the
 non-adaptive case and $\lambda_C$ is the threshold when isolation
 of symptomatic individuals and CT are activated. We consider the empirical $\Psi(s)$ during the partial opening period and the three CT strategies. The three horizontal lines are:
 the ratio $\lambda_C^{opening}/\lambda_C^{NA} \approx 2.04$ when only
 symptomatic individuals are isolated during the partial opening period (dotted);
 the value of $R_0=\lambda/\lambda_C^{NA}=4.5$ for a variant
 of SARS-CoV-2 (dashed); the ratio
 $\lambda_C^{closure}/\lambda_C^{NA} \approx 2.63 \lambda_C^{opening}/\lambda_C^{NA} \approx 5.35$ when only symptomatic individuals are isolated during the closure period (dot-dashed).
 In the inset it is plotted $s^*$ as a function of
 $\epsilon^*$ for the two targeted tracing strategies. {\bf c} We
 plot the ratio $\lambda_C/\lambda_C^{NA}$ as a function of
 $\epsilon^*$ for symptomatic isolation and activating
 progressively all CT mechanisms, considering $\Psi(s) \sim s^{-(\nu+1)}$, with $s \in
 [2,200]$ and $\nu=1.5$ and implementing a tracing
 strategy targeted at large simplices. {\bf d}
 Same of panel {\bf c} with the empirical distribution $\Psi(s)$
 during the partial opening and implementing a uniform tracing
 strategy. {\bf e} Same of panel {\bf d}
 implementing a tracing strategy targeted at small simplices. {\bf f} Same of panel {\bf d}
 implementing a tracing strategy targeted at large simplices. In all panels the parameters are fixed as discussed in Methods.}
 \label{fig:figure4}
\end{figure*}

In the presence of large simplices, epidemic diffusion is
mainly driven by SSEs, with many infections in few large gatherings.
This suggests to focus tracing efforts specifically on large simplices.
We quantitatively test this hypothesis by comparing the impact of the
different tracing mechanisms in a realistic situation.
We use $\Psi(s)$ distributions measured empirically at the University
of Parma (Italy) by considering simultaneous connections
to Access Points (APs) of the university WIFI network as proxies of gatherings (see Methods,
Fig.~\ref{fig:figure4}\textbf{a} and Fig.~\ref{fig:figure5}).
Only epidemiologically
significant gatherings, i.e. lasting longer than 15 minutes~\cite{ECDC_ct4}, are considered: note that indoor airborne transmission of SARS-CoV-2 has been observed to
occur on spatial scales comparable to those covered by
WIFI APs~\cite{Gonzalez2021WIFI}, making WIFI a good tool to support
traditional CT~\cite{Harvard2020}.
Two separate distributions $\Psi(s)$ (see Fig.~\ref{fig:figure4}\textbf{a})
are calculated for time periods
with two different levels of activity restrictions:
partial opening and closure of the University~\cite{unipr_rules} (see Methods).

Both are heterogeneous, as observed in other datasets~\cite{sekara2016fundamental}.
The
distribution in the closure period features a reduced upper cut-off,
due to the restrictions on activities involving many people,
such as in-person classes.
Consistently, the probability of simplices of size $s=0,1$
(i.e. 0 or 1 individual connected to the AP)
is significantly increased.

The variation in $\Psi(s)$ has a strong impact on the epidemic threshold.
Assuming that only isolation of symptomatic individuals is active in the
partial opening period, the epidemic threshold increases, due to the closure,
by a factor
\begin{equation}
\frac{\lambda_C^{closure}}{\lambda_C^{opening}}=\frac{\langle s(s-1)
 \rangle_{opening}}{\langle s(s-1) \rangle_{closure}} \simeq 2.63,
\end{equation}
so that $\lambda_C^{closure}/\lambda_C^{NA} \simeq 5.35$.
This change comes at the cost of stopping all teaching and most
working activities. CT aims at keeping activities
unchanged while still controlling the epidemic;
we estimate the effect of the CT strategies
during the partial opening period
and compare the mitigation due to tracing with that due to closure.

Tracing strategies are modelled through different $\epsilon(s)$.
The case $\epsilon(s)=\theta(s-s^*)$, with $\theta(x)$ the Heaviside
step function, represents a tracing targeted at large simplices, 
with size $s \geq s^*$, while smaller gatherings are not traced.
This strategy is compared with the uniform tracing adopted in
the previous section and with tracing applied only to small
simplices, $\epsilon(s)=\theta(s^*-s)$.
In the comparison, we keep constant the resources allocated for tracing,
i.e. the average fraction of traced nodes
$\epsilon^*=\frac{\langle \epsilon(s) (s-1) \rangle}{\langle s-1 \rangle}$
(in the uniform case $\epsilon(s)=\epsilon^*$ $\forall s$,
while in targeted strategies $\epsilon^*$ determines the size $s^*$).
As shown in Fig.~\ref{fig:figure4}\textbf{b}, tracing targeted at large
simplices is most effective, since it is sufficient to
trace on average a small fraction of nodes $\epsilon^*$ to obtain a
significant increase of the threshold. On the contrary,
tracing only small simplices requires almost all gatherings to be
traced to obtain a comparable result.
Tracing targeted at large simplices produces
the same effects as University closure if $\epsilon^* \gtrsim 0.47$
(see Fig.~\ref{fig:figure4}\textbf{b}), that is tracing at least all simplices
with $s \geq 6$, representing $16.1 \%$ of simplices with $s \geq 2$.
This is a quite demanding goal.
However, closures are measures which may be more
drastic than what strictly needed. For example,
the University closure, increasing the threshold to
$\lambda_C^{closure}/\lambda_C^{NA} \simeq 5.35$
is somehow an overreaction against
a variant of SARS-CoV-2 with a basic reproduction number
$R_0=\lambda/\lambda_C^{NA} \approx 4.5$ (e.g. a highly transmissible variant in the presence of additional mitigation measures such as the use of face masks and physical distancing)~\cite{who_china_report,DiDomenico2020,who_variants,davies2021estimated,Volz2021,imperial2021variants}.
The epidemic can instead be kept under control by
allowing a partial opening of University activities
while tracing large gatherings.

As shown in Fig.~\ref{fig:figure4}\textbf{b}, for remaining below the
threshold it is enough to trace on average a reasonable fraction
of nodes per simplex $\epsilon^* \gtrsim 0.27$, corresponding to
tracing all simplices with $s \geq 9$,
which represent only the $6.2 \%$ of simplices with $s \geq 2$.

The bottom panels of Fig.~\ref{fig:figure4} describe the different
contributions of forward, backward and sideward CT for the various
strategies. Clearly backward tracing provides the most relevant
contribution to the threshold increase.
However, in the strategy targeted at large simplices, a relevant contribution is also
provided by sideward CT, which turns out to be a fundamental ingredient
when large simplices sustain epidemic transmission. Note however
that in the period of partial opening several restrictions are still
in place (e.g. the maximum capacity of a classroom is reduced by a factor $4$~\cite{unipr_rules}):
this explains the relatively small cut-off $s_M \approx 60$
in the $\Psi(s)$ distribution. For broader distributions with larger cut-off
(corresponding to full opening of the University)
sideward tracing is expected to provide the dominant contribution, as shown
in Fig.~\ref{fig:figure4}\textbf{c}, where we plot the different contributions to the
targeted tracing for a power-law distribution
$\Psi(s) \sim s^{-(\nu+1)}$, with $\nu=1.5$ and a cut-off $s_M=200$.

In the SM, we check the 
stability of the results under small variations of the parameters
for the empirical and the synthetic distributions.
In particular we consider the case where in the targeted strategy
large clusters are not perfectly identified.

\section{Discussion}

Tracing potentially infectious people is one of the fundamental
non-pharmaceutical interventions to control epidemic spreading.
This is even more true now, as in many countries new extensive outbreaks are emerging due to highly transmissible SARS-CoV-2 variants, and new restrictions are implemented.
In this paper we have pointed out that tracing all participants in a
gathering augments usual backward and forward tracings with a new
dimension, sideward tracing, allowing to detect transmission
events that would be missed otherwise. Our analysis indicates that this
may lead to a large improvement of the efficacy of CT in many
situations, especially if targeted at large simplices, as shown explicitly with data on real gatherings at a
University Campus during the current pandemic.
Our model does not include several ingredients
that may increase the accuracy of the description of real
systems, such as the presence of immunised people, 
the different infectivity of symptomatic and asymptomatic individuals, the dependence of the probability of infection on the simplex size, the
existence of delays associated to the tracing process 
or the different probability of tracing individuals within the same simplex.
Nevertheless, we expect these effects not to change the overall
picture and the relevance of sideward tracing.
Let us also point out that we have considered only the increase
of the epidemic threshold as a measure of CT efficacy.
Another central observable is the prevalence for values
of $\lambda$ above the transition. For such a quantity we expect
the effect of sideward tracing to be even larger, given the
nonlinear effects due to the presence of more than one infected
individual in a single gathering.
Let us also stress the relevance of the application of our
approach to a specific empirical case, where gatherings are reconstructed
based on connections to a WIFI network. This type of passively obtained
data constitutes a potential trove of information for tracing purposes.
Many large organisations such as universities, schools, companies,
already collect these data for their normal functioning.
With suitable privacy-preserving protocols to treat them,
they may provide an extremely simple and cheap way to trace gatherings
as a whole and thus add a new weapon in the battle against disease spreading.

\section{Methods}

\subsection{Mean-field equations and epidemic threshold}
The epidemic threshold is obtained from the linearisation of the evolution equations. For activity-driven models, the results of the compartmental model are exact since the random selection of participants in 
the gatherings destroys local correlations. Clearly, in an 
intrinsically mean-field model finite size effects could be present, 
however vanishing as the number of agents grows. Moreover, 
in this framework
the thresholds for SIR and SIS models coincide~\cite{Tizzani2018}.
Therefore we consider the slightly simpler SIS dynamics and
we write down equations for the temporal evolution
(for arbitrary $\Psi(s)$ and $\epsilon(s)$) of the probabilities
that at time $t$ a node is in each of the epidemiological
compartments, i.e. $S(t)$,
$A(t)$, $A_T(t)$, $A_Q(t)$, $P(t)$ and $I(t)$.
We obtain a set of coupled non-linear differential
equations which describe the agents activity, the epidemic spreading
and the adaptive behaviour due to isolation of the symptomatic and CT
(see SM for the equations and their derivation).
By varying the control parameter $\lambda$, the model features a
transition between a phase where outbreaks involve only a finite number
of individuals and a phase where they extend to a finite fraction
of the population.
Linearising around the healthy phase $S=1$, we obtain
\begin{widetext}
\begin{equation}
 \left \{
 \begin{array}{lll}
 \partial_t P(t) &=& -\gamma_P P(t) + \lambda \delta a \langle s(s-1)\rangle [A(t)+A_T(t)+P(t)] \\
 \partial_t I(t) &=& -\mu_I I(t) + \gamma_P P(t) \\
 \partial_t A(t) &=& -\mu A(t) + \lambda (1-\delta) a \langle s(s-1)\rangle [A(t)+A_T(t)+P(t)] \\
 & & - C^{Forward} - C^{Backward} - C^{Sideward} \\
 \partial_t A_T(t) &=& -(\mu+\gamma_P) A_T(t) + C^{Forward} + C^{Backward} + C^{Sideward} \\
 \partial_t A_Q(t) &=& -\mu A_Q(t) + \gamma_P A_T(t) 
 \end{array}
 \right .
\label{linearised}
\end{equation}
\end{widetext}
where
\begin{equation}
 C^{Forward} = \lambda (1-\delta) a \langle \epsilon(s) s (s-1) \rangle P(t),
 \label{eq:CTfor}
\end{equation}
\begin{equation}
 C^{Backward} = a \langle \epsilon(s) s \left[1- (1-\lambda \delta)^{s-1} \right]
 \rangle A(t),
 \label{eq:CTback}
\end{equation}
while $C^{Sideward}$ is given by Eq.~\eqref{eq:CTside}.
In Eq.~\eqref{eq:CTfor} $a$ is the simplex activation rate, the
probability to take part to the simplex for the susceptible node is
proportional to $s$ and, in the linearised regime, the probability that among the remaining $(s-1)$ nodes there is a presymptomatic node is $(s-1)P(t)$. The asymptomatic infection occurs with probability $\lambda(1-\delta)$, $\epsilon(s)$ is the probability to trace the simplex and the term is averaged over simplex size.
In Eq.~\eqref{eq:CTback} $a$ is the simplex activation rate, the
probability to take part to the simplex for the asymptomatic node $A$ is
proportional to $s$ and $\left[ 1- (1-\lambda \delta)^{(s-1)}\right]$ is the probability that
at least one of the remaining $(s-1)$ nodes is infected and
symptomatic. Again $\epsilon(s)$ is the probability that the simplex
is traced and the term is averaged over simplex size.

The solution $S=1$ becomes unstable for $\lambda$ above the threshold $\lambda_C$,
which can be calculated from the largest eigenvalue of the Jacobian Matrix
associated to Eq.~\eqref{linearised}. It can be determined analytically
(in some specific cases) and numerically (in general).
See SM for the detailed derivations.

\begin{figure*}
\centering
\includegraphics[width=180mm]{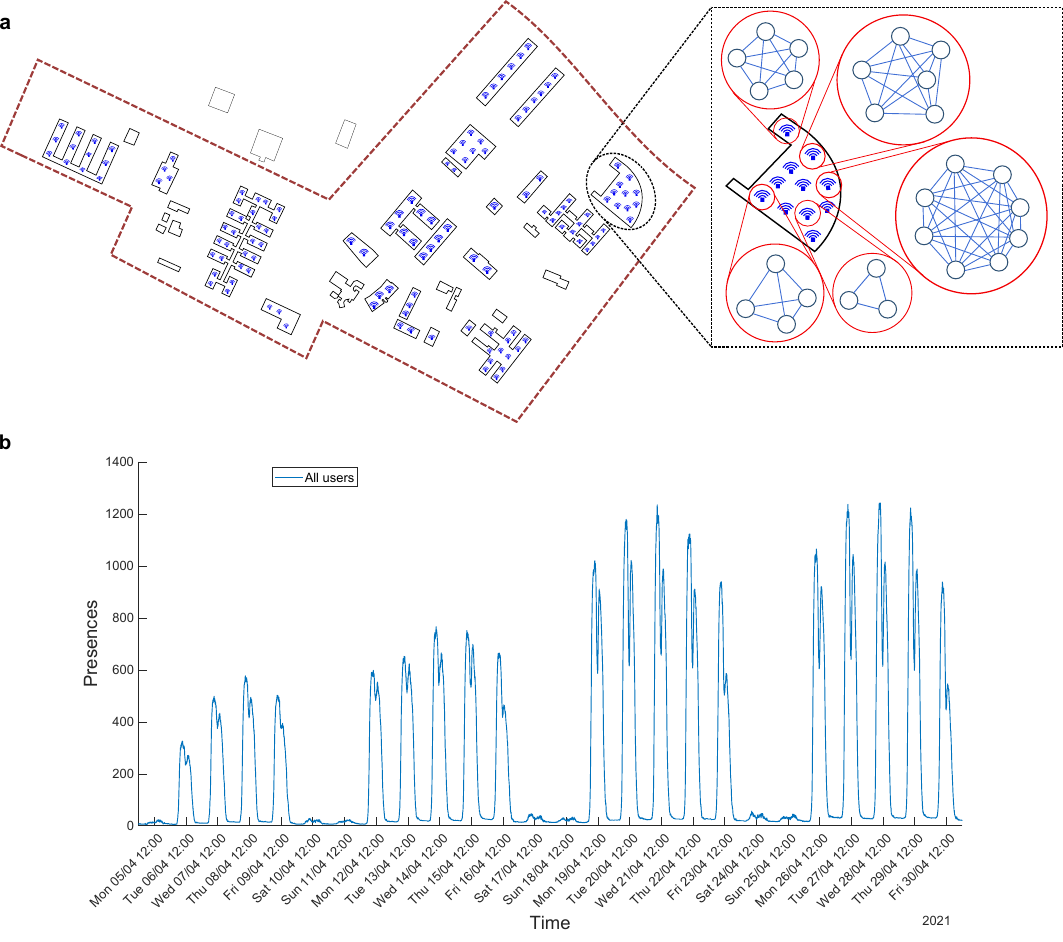}
 \caption{{\bf WIFI data for gatherings at University of Parma.} {\bf
 a} Map of the University of Parma Campus, called "Parco
 Area delle Scienze", where scientific departments are located
 with facilities for teaching and research activities (classrooms,
 laboratories, study spaces and libraries). The map highlights
 schematically the buildings where the gatherings are recorded
 through WIFI access points data. {\bf b} Plot of the temporal
 dependence of the number of users connected to WIFI from 05th April
 2021 to 30th April 2021.
 On Saturdays and Sundays the accesses were
 minimal, as well as on 05th April 2021 being Easter Monday (public
 holiday in Italy). A rapid increase in the number of users occurred in the
 morning, followed by a reduction in connections at lunchtime, a
 resumption in the afternoon and a dramatic decrease in the evening. The
 plot covers both partial opening and closure periods due
 to COVID-19 containment measures: from 05th April to 18th April
 (closure period) attendance was low, while from
 19th April to 30th April (partial opening period) it increased
 significantly.}
 \label{fig:figure5}
\end{figure*} 

\subsection{Model parameters}
Throughout the paper the parameter values are, unless
specified otherwise, tailored to describe the SARS-CoV-2
diffusion. The probability for an individual to develop symptoms is
$\delta=0.57$~\cite{Lavezzo2020}. The average time for the appearance of symptoms,
i.e. the average time after which a presymptomatic
individual spontaneously develops symptoms, is $\tau_P =1/\gamma_P=1.5$
days~\cite{DiDomenico2020,wei2020presymptomatic}. The
average recovery time is $\tau=1/\mu=14$
days~\cite{who_china_report,Liu2020}. In order to implement all CT
mechanisms, we assume contacts are reconstructed over $T_{CT} = \tau =
14$ days, so that it is possible to track nodes
infected by the index case in its presymptomatic phase (forward CT),
the source of infection of the index case (backward CT) and all other
infections occurred in simplicial interactions (sideward CT). The
average number of daily contacts per individual is fixed to
$\overline{n}=14$~\cite{Mossong2008, Vankerkove2013}.
 Notice that with these realistic values of the parameters, $R_0\approx 4.5$ corresponds to $\lambda \approx 0.02$. This implies that even in the largest gathering that we consider of about hundreds of nodes, at most only 10-20 individuals can be infected.

\subsection{WIFI data: preprocessing and main properties}
Like many universities, the University of Parma has covered its
buildings with a single WIFI network (see Fig.~\ref{fig:figure5}\textbf{a}),
enabling users to establish more than 10000 sessions a day. All
sessions data from the login management system are collected by the "ICT services" office of Parma University. The login management system manages all
wireless access points (APs) and all users' requests for connection to
the internet with their registered device. University WIFI users are
all people with institutional account (students, professors,
researchers and staff) and all people from other European Universities
with an account for the EDUROAM network. It is possible to create temporary
accounts for external guests.

The dataset refers to a sample of 713 wireless access points, about 10000 daily connections and spans six months, starting on 10th December 2020 and ending on 09th May 2021. During this period, due to the
COVID-19 pandemic restrictions, we can distinguish two different
phases: a closure phase and a partial opening phase~\cite{unipr_rules}.

\begin{itemize}
 \item \textbf{Closure phase.} During this phase, the access to all
 University buildings was allowed only to authorised staff,
 professors, researchers and students only to participate in
 laboratory activities. All lectures were offered remotely on
 video-conferencing platforms. This phase spans the period
 from 10th December 2020 to 21st February 2021 and from 15th
 March 2021 to 18th April 2021. In this period the number of users is 7138;
 
 \item \textbf{Partial opening phase.} During this phase, the access
 to University buildings was extended to first year students
 attending in-person classes (about 25\% students
 enrolled in the degree courses) while the lectures of all others students were offered remotely on video-conferencing platforms. Libraries and study spaces reopened with a reduction of rooms' maximum capacity and mandatory reservation. This phase spans the
 period from 22nd February 2021 to 14th March 2021 and from 19th April
 2021 to 09th May 2021. In this period the number of users is 7835;
\end{itemize}

We provided the "ICT services" office with a script that extracts from the login management system the anonimysed aggregated data
on the attendance at the University and its temporal dynamics (see SM). In Fig.~\ref{fig:figure5}\textbf{b} we plot the number of people
present at the University as a function of time. Data show the
reliability of the measure: the closure period up to the 18th of April
and the following days of partial opening are clearly distinct on the
plot; in the week-ends and on public holidays (April 5th) attendance
is limited; during night hours connections are nearly
absent and we also observe the effect of
lunch breaks. The attendance estimated via WIFI data underestimates
the true number of people simultaneously present, since not all individuals
connect to the WIFI network.
A reasonable estimate is that the number of people actually present
is roughly twice as large.
This assessment is based on the comparison of WIFI data with
the number of seat reservations in classrooms and study rooms made by students
(mandatory for University regulations~\cite{unipr_rules}) and
with the number of attending professors,
researchers and university staff, as revealed by data from their
personal badges.

\subsection{Empirical simplex size distribution from WIFI data}
From the WIFI dataset of the University of Parma, we extract two different
simplex size distributions: one for the closure phase and one for
the partial opening phase. To control the signal noise of day-to-day
random variations, we take the entire dataset and remove public
holidays, weekends and the night time data (from 8:00 p.m.
to 7:00 a.m.). We define as a
simplex of size $s$ a group of $s$ users connected to the same AP for
at least 15 minutes (the same time used in contact tracing apps to
consider a contact epidemiologically significant and at high-risk~\cite{ECDC_ct4}). For both
phases, we reconstruct all connection sessions of all users,
their location and their duration from the WIFI log file.
It is possible that
some connections are interrupted for various technical reasons (such as weak
signal or the user's device going in standby): to control for these effects,
if a user connects twice to the same AP and the time lag $\Delta t$
between the two sessions is smaller than 5 minutes, we consider it as a single connection.
To obtain all simplices activated inside the University we focus
on connections to a given AP.
We split working hours (from 7:00
a.m. to 8:00 p.m.) into 15 minutes intervals and for each interval we
find the number of users that remained connected to the same AP
for the full time interval: this number corresponds to the simplex size.
This operation is repeated for every working day and for each AP. These data,
separated for the two periods, determine the empirical simplex size distributions
(Fig.~\ref{fig:figure4}\textbf{a}).

\section{Authors' contributions}
M.M., A.G., C.C., A.V. and R.B. designed research, M.M., A.G., C.C., A.V. and R.B. performed research, M.M. and A.G. analyzed data, M.M., A.G., C.C., A.V. and R.B. wrote the paper.

\section{Competing interests}
The authors declare no competing interests.

\section{Acknowledgements}
We warmly thank the Operative Unit "Technological Systems and Infrastructures" of the University of Parma and in particular A. Barontini, F. Russo and C. Valenti for the constant support in data elaboration. We also thank M. Mamei for very useful discussions and suggestions.

\section{Funding}
R.B., A.G. and A.V. acknowledge support from the EU-PON Project POR FSE Emilia Romagna 2018/2020 Objective 10.


\end{document}


\title{Supplementary Material for "Sideward contact tracing and the control of epidemics in large gatherings"}

\author{Marco Mancastroppa}
\affiliation{Dipartimento di Scienze Matematiche, Fisiche e Informatiche,
 Universit\`a degli Studi di Parma, Parco Area delle Scienze, 7/A 43124 Parma, Italy}
\affiliation{INFN, Sezione di Milano Bicocca, Gruppo Collegato di Parma, Parco Area delle Scienze, 7/A 43124 Parma, Italy}
\author{Andrea Guizzo}
\affiliation{Dipartimento di Scienze Matematiche, Fisiche e Informatiche,
 Universit\`a degli Studi di Parma, Parco Area delle Scienze, 7/A 43124 Parma, Italy}
\affiliation{INFN, Sezione di Milano Bicocca, Gruppo Collegato di Parma, Parco Area delle Scienze, 7/A 43124 Parma, Italy}
\author{Claudio Castellano} 
\affiliation{Istituto dei Sistemi Complessi (ISC-CNR), Via dei Taurini 19, I-00185 Roma, Italy}
\author{Alessandro Vezzani}
\affiliation{Istituto dei Materiali per l'Elettronica ed il Magnetismo (IMEM-CNR), Parco Area delle Scienze, 37/A-43124 Parma, Italy}
\affiliation{Dipartimento di Scienze Matematiche, Fisiche e Informatiche,
 Universit\`a degli Studi di Parma, Parco Area delle Scienze, 7/A 43124 Parma, Italy}
\affiliation{INFN, Sezione di Milano Bicocca, Gruppo Collegato di Parma, Parco Area delle Scienze, 7/A 43124 Parma, Italy}
\author{Raffaella Burioni}
\affiliation{Dipartimento di Scienze Matematiche, Fisiche e Informatiche,
 Universit\`a degli Studi di Parma, Parco Area delle Scienze, 7/A 43124 Parma, Italy}
\affiliation{INFN, Sezione di Milano Bicocca, Gruppo Collegato di Parma, Parco Area delle Scienze, 7/A 43124 Parma, Italy}

\maketitle

\tableofcontents

\bigskip

In this Supplementary Material we present the detailed derivation of the mean-field equations for an epidemic process on adaptive simplicial activity-driven networks, in the presence of contact tracing (CT) within simplices. We present the evaluation of the epidemic threshold and we discuss in detail the stability of the results with respect to the model parameters. We also investigate the effects of a fraction of large simplices which are not traced in the targeted CT strategy, due to realistic errors in CT. Finally, we describe in detail the WIFI data and their analysis.

\section{Supplementary Method 1: Mean-field equations and derivation of the epidemic threshold} 
We consider the epidemic model proposed in the main text, evolving on an adaptive activity-driven network in the presence of simplicial interactions. The containment measures are implemented as contact tracing of asymptomatic nodes with its forward, backward and sideward implementations. The epidemic model proposed is a Susceptible-Infected-Recovered (SIR) model, with further distinctions for the stage of the infection. The distinctions are based on the presence of symptoms ($P$ - infected presymptomatic, $I$ - infected symptomatic, $A$ - infected asymptomatic), on tracing and isolation ($A_T$ - asymptomatic traced and $A_Q$ - asymptomatic quarantined) and they model changes in social behaviour depending on nodes' health status. We do not consider here memory~\cite{Tizzani2018} nor burstiness effects in the dynamics~\cite{Mancastroppa2019}. The allowed transitions among these states are shown in Supplementary Fig.~\ref{fig:figure1}, which coincides with Fig. 1\textbf{b} of the main text and which we reproduce here for simplicity.

\begin{figure}
 \includegraphics[width=0.6\textwidth]{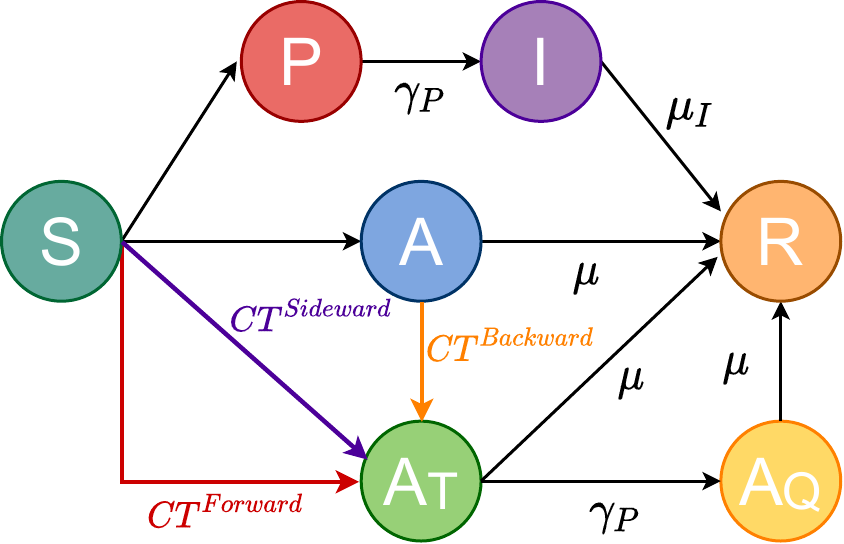}
 \caption{\textbf{Epidemic model with contact tracing.} We plot the scheme of the compartmental epidemic model together with the transitions for CT. The rates for the infection events and for the CT mechanisms are not indicated here but they are detailed in the text.}
 \label{fig:figure1}
\end{figure}

The infection and CT events are discussed in the main text, together with the CT mechanisms. An infection can be symptomatic with probability $\delta$ or asymptomatic with probability $(1-\delta)$. We consider the manual implementation of CT~\cite{mancastroppa2021contacttracing}: the tracing is performed on the whole simplex in which the symptomatic individual participated, tracing all nodes engaged in the gathering, and the tracing is effective with probability $\epsilon(s)$, with $s$ the simplex size. 

The infection and tracing transitions are described below by considering the necessary and sufficient (minimum) conditions for them to happen in a simplex.
\begin{align}
P+S& \xrightarrow[]{\lambda \delta} P+P \label{eq:1}\\
P+S& \xrightarrow[]{\lambda (1-\delta) \epsilon} P+A_T \label{eq:2}\\
P+S& \xrightarrow[]{\lambda (1-\delta) (1-\epsilon)} P+A \label{eq:3}
\end{align}
A susceptible node $S$ can be infected by a presymptomatic node $P$: if the infection is symptomatic the susceptible node becomes presymptomatic $P$ (Eq.~\eqref{eq:1}); if the infection is asymptomatic, the contagion can be traced by forward CT activated by the infector $P$ and the susceptible node becomes traced asymptomatic $A_T$ (Eq.~\eqref{eq:2}), otherwise if the simplex is not traced the susceptible node becomes asymptomatic $A$ (Eq.~\eqref{eq:3}).

\begin{align}
&A+S \xrightarrow[]{\lambda \delta \epsilon} A_T+P \label{eq:4}\\
&A+S \xrightarrow[]{\lambda \delta (1-\epsilon)} A+P \label{eq:5}\\
&A_T+S \xrightarrow[]{\lambda \delta} A_T+P \label{eq:6}
\end{align}
A susceptible node $S$ can be infected by an asymptomatic $A$ (or traced asymptomatic $A_T$) node with symptomatic infection and thus becomes presymptomatic $P$. The asymptomatic infector can be traced by backward CT, activated by the newly infected $P$, becoming traced asymptomatic $A_T$ (Eq.~\eqref{eq:4}), otherwise if the simplex is not traced the infector does not change status (Eq.~\eqref{eq:5}). In principle also a traced asymptomatic infector can be traced with backward CT but this has actually no consequences (Eq.~\eqref{eq:6}). In the event of Eq.~\eqref{eq:4} both individuals change state.

\begin{align}
&A+S \xrightarrow[]{\lambda (1-\delta) \epsilon} A+A_T \qquad &\text{if in the same simplex exists}& \qquad A+S \xrightarrow[]{\lambda \delta} A+P
\label{eq:A1}\\
&A+S \xrightarrow[]{\lambda (1-\delta) (1-\epsilon)} A+A \qquad &\text{if in the same simplex exists}& \qquad A+S \xrightarrow[]{\lambda \delta} A+P 
\label{eq:A2}\\
&A+S \xrightarrow[]{\lambda (1-\delta)} A+A \qquad &\text{otherwise} \label{eq:A3}\\
&A_T+S \xrightarrow[]{\lambda (1-\delta) \epsilon} A_T+A_T \qquad &\text{if in the same simplex exists}& \qquad A_T+S \xrightarrow[]{\lambda \delta} A_T+P
\label{eq:AT1}\\
&A_T+S \xrightarrow[]{\lambda (1-\delta) (1-\epsilon)} A_T+A \qquad &\text{if in the same simplex exists}& \qquad A_T+S \xrightarrow[]{\lambda \delta} A_T+P 
\label{eq:AT2}\\
&A_T+S \xrightarrow[]{\lambda (1-\delta)} A_T+A \qquad &\text{otherwise} \label{eq:AT3}
\end{align}
A susceptible node $S$ can be infected by an asymptomatic $A$ (or traced asymptomatic $A_T$) node with asymptomatic infection. If in the same simplex at least a symptomatic infection occurs, CT can be activated on the simplex by the newly infected $P$. The asymptomatic contagion can be traced by sideward CT and the susceptible node becomes traced asymptomatic $A_T$ (Eqs.~\eqref{eq:A1}-\eqref{eq:AT1}), otherwise the susceptible node becomes infected asymptomatic $A$ if the simplex is not traced (Eqs.~\eqref{eq:A2}-\eqref{eq:AT2}) or if no symptomatic infections occur in the same simplex (Eqs.~\eqref{eq:A3}-\eqref{eq:AT3}). In the events of Eqs.~\eqref{eq:AT1}-\eqref{eq:AT2} the infector $A_T$ can be in principle traced with backward CT but this has actually no consequences; in the events of Eqs.~\eqref{eq:A1}-\eqref{eq:A2} the infector $A$ can also be traced with backward tracing as indicated in Eqs.~\eqref{eq:4}-\eqref{eq:5}.

Finally, the spontaneous transitions are:
\begin{align}
&P \xrightarrow[]{\gamma_P} I& \qquad \qquad \qquad &A \xrightarrow[]{\mu} R& \qquad \qquad \qquad &I \xrightarrow[]{\mu_I} R&\\
&A_T \xrightarrow[]{\gamma_P} A_Q& \qquad \qquad \qquad &A_T \xrightarrow[]{\mu} R& \qquad \qquad \qquad &A_Q \xrightarrow[]{\mu} R&
\end{align}
which account for spontaneous recovery, spontaneous symptoms development and isolation of traced individuals.

We apply a \textit{mean-field} approach, which is exact since
local correlations are
continuously destroyed because of link reshuffling. Therefore,
the epidemic thresholds of the $SIR$ and $SIS$
(Susceptible-Infected-Susceptible) models
coincide~\cite{Tizzani2018}. This allows us to determine the
threshold by considering the
mean-field equations for the $SIS$ version of the dynamics.

We consider the probability for a node to belong to a compartment of the epidemic model shown in Supplementary Fig.~\ref{fig:figure1} and we derive the equations for the temporal evolution of the probabilities, taking into account the network dynamics, the epidemic process and the CT dynamics. At the mean-field level, the epidemic dynamics is described by the probabilities:
\begin{itemize}
\item $P(t)$ for a node to be infected presymptomatic at time $t$;
\item $I(t)$ for a node to be infected symptomatic at time $t$;
\item $A(t)$ for a node to be infected asymptomatic at time $t$;
\item $A_T(t)$ for a node to be asymptomatic traced at time $t$;
\item $A_Q(t)$ for a node to be asymptomatic isolated at time $t$;
\item $S(t)=1-P(t)-I(t)-A(t)-A_T(t)-A_Q(t)$ for a node to be susceptible at time $t$.
\end{itemize}

The network evolves as follows: simplices activate with a Poissonian dynamics with activation rate $a$ and at each activation their size is drawn from the distribution $\Psi(s)$. Nodes are assigned with an attractiveness parameter $b_i$, which tunes the propensity of an individual to engage social interactions: thus nodes participate in active simplices with probability $p_{b_i} \propto b_i$~\cite{Pozzana2017,Mancastroppa2020,mancastroppa2021contacttracing}.
We consider all nodes with the same attractiveness ($b_S=b$) when susceptible, so they participate equally in active simplices. Presymptomatic, asymptomatic and traced asymptomatic, behave as if they were susceptible
($b_P=b_A=b_{A_T}=b_S=b$), while symptomatic and quarantined asymptomatic are isolated and do not participate in simplices ($b_I=b_{A_Q}=0$). In this case the average attractiveness at time $t$ is $\overline{b}(t) = b(S(t)+P(t)+A(t)+A_T(t))$. We consider the thermodynamic limit.
The probability $P(t)$ that a node is in the presymptomatic state evolves according to the following equation:

\begin{equation}
\begin{aligned}
\partial_t P(t)= -\gamma_P P(t) + \int ds \Psi(s) a s P_S(t) Z_s(t) \delta
\end{aligned}
\end{equation}
where the first term on the right hand side accounts for symptoms
development and the second term accounts for contagion processes. In
particular, the latter is given by the product of the activation rate $a$
of a simplex of size $s$, the probability $s P_S(t)$ that one susceptible
node participates in it, the probability $Z_s(t)$ that at least one of the other
$(s-1)$ nodes infects the susceptible one and the probability $\delta$ that the
infection is symptomatic. Both terms are averaged over the size of the
simplex $\int ds \Psi(s)$.

Let us denote as $P_X(t)$ the probability that a node in an active simplex
belongs to the compartment $X$. Thus, by definition:
\begin{equation}
P_X(t) = X(t) \frac{b_X}{\overline{b}(t)}
\end{equation}
Therefore $P_I(t)=P_{A_Q}(t)=0 \, \forall t$ and $P_X(t)= \frac{X(t)}{S(t)+P(t)+A(t)+A_T(t)} \,$ for $X=S,P,A,A_T$.

The probability $Z_s(t)$ that at least one of the other $(s-1)$ nodes
of the simplex infects the susceptible one is
\begin{equation}
Z_s(t)=1-\xi(t)^{s-1}
\end{equation}

where $\xi(t)$ is the probability that a node in the simplex does not infect the susceptible one. This quantity can be written as
\begin{equation}
\xi(t)=P_S(t)+(1-\lambda)[P_P(t)+P_A(t)+P_{A_T}(t)]=1-\lambda\frac{P(t)+A(t)+A_T(t)}{S(t)+P(t)+A(t)+A_T(t)}
\end{equation}
Thus the complete equation for the evolution of $P(t)$ is:
\begin{equation}
\begin{aligned}
\partial_t P(t)= -\gamma_P P(t) + a \frac{S(t)}{S(t)+P(t)+A(t)+A_T(t)} \delta \left\langle s \left[1-\left(1-\lambda \frac{P(t)+A(t)+A_T(t)}{S(t)+P(t)+A(t)+A_T(t)}\right)^{s-1}\right] \right\rangle
\end{aligned}
\end{equation}
where we indicate with $\langle f(s) \rangle = \int ds \Psi(s) f(s)$: the first term on the right hand side accounts for spontaneous recovery and the second term for symptomatic infections in simplices.

The equation for the probability $I(t)$ that a node is in the symptomatic
infected state is trivially
\begin{equation}
\partial_t I(t)=-\mu_I I(t) + \gamma_P P(t)
\end{equation}
where the first term on the right hand side accounts for spontaneous recovery and the second term for spontaneous symptoms development.

The equation for the probability $A(t)$ to be in the asymptomatic (untraced,
nonquarantined) state is the most complex:
\begin{equation}
\begin{aligned} 
\partial_t A(t)=-\mu A(t) &+ \int ds \Psi(s) a s P_S(t) Z_s(t) (1-\delta) \\
&- \int ds \Psi(s) C_s^{Forward}(t)\\
&- \int ds \Psi(s) C_s^{Backward}(t)\\
&- \int ds \Psi(s) C_s^{Sideward}(t).
\end{aligned}
\end{equation}
The first term on right hand side accounts for spontaneous recovery;
the second term accounts for contagion processes; the third, fourth
and fifth terms account respectively for forward, backward and
sideward CT. All terms are averaged over the size of the simplex
$\int ds \Psi(s)$. The infection term accounts for the activation $a$ of a
simplex of size $s$, the probability that one susceptible node
participates in it $s P_S(t)$ and that at least one of the other
$(s-1)$ nodes infects the susceptible one $Z_s(t)$ with an
asymptomatic infection $(1-\delta)$. Its evaluation is analogous to
the one described for $P(t)$. Hereafter we evaluate separately each CT
term $C^L= \int ds \Psi(s) C_s^L(t)$, with $L=Forward,Backward,Sideward$.

\subsection{Forward CT}
\begin{equation}
C^{Forward}=\int ds \Psi(s) a sP_S(t) \epsilon(s) (1-\delta) F_s(t) 
\end{equation}
The forward CT, described in details in the main text, traces an
asymptomatic individual infected by a presymptomatic node which will
develop symptoms and activate CT. Thus, the forward CT term accounts
for the activation $a$ of a simplex of size $s$, for the probability
$sP_S(t)$ that a susceptible node participates in the simplex
and the probability $F_s(t)$ that at least one of the other $(s-1)$ nodes is
a presymptomatic node $P$ which infects the susceptible one with an
asymptomatic infection $(1-\delta)$. The simplex is traced with
probability $\epsilon(s)$.
\begin{equation}
F_s(t)=1-k(t)^{s-1}
\end{equation}
is the probability that at least one of the other $(s-1)$ nodes is
presymptomatic and infects the susceptible one and thus $k(t)$ is the
probability that a node in the simplex is not $P$ or if he/she is
presymptomatic he/she does not infect the susceptible node.
\begin{equation}
k(t)=P_S(t)+P_A(t)+P_{A_T}(t)+(1-\lambda)P_P(t)=1-\lambda \frac{P(t)}{S(t)+P(t)+A(t)+A_T(t)}
\end{equation}
Thus, we obtain:
\begin{equation}
C^{Forward}= a \frac{S(t)}{S(t)+P(t)+A(t)+A_T(t)}(1-\delta) \left\langle \epsilon(s) s \left[1-\left(1-\lambda \frac{P(t)}{S(t)+P(t)+A(t)+A_T(t)}\right)^{s-1}\right]\right\rangle
\end{equation}

\subsection{Backward CT}
\begin{equation}
C^{Backward}=\int ds \Psi(s) a sP_A(t) \epsilon(s) W_s(t)
\end{equation}
The backward CT, described in details in the main text, traces an
asymptomatic individual who infects a susceptible node producing a
symptomatic infection. Thus, the backward CT term accounts for the
activation $a$ of a simplex of size $s$, for the probability $sP_A(t)$
that an asymptomatic node participates in the simplex 
and the
probability $W_s(t)$ that at least one of the other $(s-1)$ nodes is
a susceptible node which is infected by the asymptomatic one with a
symptomatic infection. The simplex is traced with probability
$\epsilon(s)$.
\begin{equation}
W_s(t)=1-\phi(t)^{s-1}
\end{equation}
is the probability that at least one of the other $(s-1)$ nodes is a susceptible node which is infected by the asymptomatic one with a symptomatic infection and thus $\phi(t)$ is the probability that a node in the simplex is not a susceptible node infected with presymptomatic infection. 
\begin{equation}
\phi(t)=P_P(t)+P_A(t)+P_{A_T}(t)+(1-\lambda)P_S(t)+\lambda(1-\delta)P_S(t)=1-\lambda \delta \frac{S(t)}{S(t)+P(t)+A(t)+A_T(t)}
\label{phi}
\end{equation}
Thus, we obtain:
\begin{equation}
C^{Backward}= a \frac{A(t)}{S(t)+P(t)+A(t)+A_T(t)} \left\langle \epsilon(s) s \left[1-\left(1-\lambda \delta \frac{S(t)}{S(t)+P(t)+A(t)+A_T(t)}\right)^{s-1}\right]\right\rangle
\end{equation}

\subsection{Sideward CT}
\begin{equation}
C^{Sideward}=\int ds \Psi(s) a sP_S(t) (1-\delta) H_s(t) \epsilon(s) K_s(t) 
\end{equation}
The sideward CT, described in details in the main text, traces an
asymptomatic individual who is infected by an asymptomatic (or traced
asymptomatic) if in the same simplex also another susceptible node is
infected with a symptomatic infection and then will activate CT.
Thus, the sideward CT term contains the activation $a$ of a
simplex of size $s$, the probability $sP_S(t)$ that a susceptible node
participates in the simplex, the probability $H_s(t)$
that at least one of the other $(s-1)$ nodes is infected asymptomatic
(or traced asymptomatic) and infects the susceptible one with an
asymptomatic infection $(1-\delta)$, the probability $K_s(t)$ that
among the remaining $(s-2)$ nodes at least one of them is a
susceptible node infected in the simplex with a symptomatic
infection. The simplex is traced with probability $\epsilon(s)$.
Analogously to the cases of forward and backward CT:
\begin{equation}
H_s(t)=1-h(t)^{s-1}
\end{equation}
where $h(t)$ is the probability for a node of the simplex not to infect the susceptible node unless he/she is $P$.
\begin{equation}
h(t)=P_S(t)+(1-\lambda)(P_A(t)+P_{A_T}(t))+P_P(t)=1-\lambda \frac{A(t)+A_T(t)}{S(t)+P(t)+A(t)+A_T(t)}
\end{equation}
Analogously
\begin{equation}
K_s(t)=1-\phi(t)^{s-2}
\end{equation}
where $\phi(t)$ is given by Eq.~\eqref{phi}.

Thus, we obtain:
\begin{equation}
C^{Sideward}= a \frac{S(t)}{S(t)+Y(t)} (1-\delta) \left\langle \epsilon(s) s \left[1-\left(1-\lambda \frac{A(t)+A_T(t)}{S(t)+Y(t)}\right)^{s-1}\right] \left[1-\left(1-\lambda \delta \frac{S(t)}{S(t)+Y(t)}\right)^{s-2}\right]\right\rangle
\end{equation}
where $Y(t)=P(t)+A(t)+A_T(t)$.\\

The complete equation for the evolution of $A(t)$ is:
\begin{equation}
\begin{aligned}
\partial_t A(t)=&-\mu A(t) + a \frac{S(t)}{S(t)+P(t)+A(t)+A_T(t)} (1-\delta) \left\langle s \left[1-\left(1-\lambda \frac{P(t)+A(t)+A_T(t)}{S(t)+P(t)+A(t)+A_T(t)}\right)^{s-1}\right]\right\rangle\\
-& a \frac{S(t)}{S(t)+P(t)+A(t)+A_T(t)}(1-\delta) \left\langle \epsilon(s) s \left[1-\left(1-\lambda \frac{P(t)}{S(t)+P(t)+A(t)+A_T(t)}\right)^{s-1}\right]\right\rangle\\ 
-& a \frac{A(t)}{S(t)+P(t)+A(t)+A_T(t)} \left\langle \epsilon(s) s \left[1-\left(1-\lambda \delta \frac{S(t)}{S(t)+P(t)+A(t)+A_T(t)}\right)^{s-1}\right]\right\rangle \\
-& a \frac{S(t)}{S(t)+Y(t)} (1-\delta) \left\langle \epsilon(s) s \left[1-\left(1-\lambda \frac{A(t)+A_T(t)}{S(t)+Y(t)}\right)^{s-1}\right]\left[1-\left(1-\lambda \delta \frac{S(t)}{S(t)+Y(t)}\right)^{s-2}\right]\right\rangle
\end{aligned}
\end{equation}
where the first term on the right hand side accounts for spontaneous recovery, the second term for asymptomatic infections in simplices and the third, fourth and fifth terms account respectively for forward, backward and sideward CT.

The equation for the probability $A_T(t)$ to be in the asymptomatic traced state is
\begin{equation}
\partial_t A_T(t)=-(\mu+\gamma_P) A_T(t) + C^{Forward} + C^{Backward} +C^{Sideward}
\end{equation}

Thus substituting:

\begin{equation}
\begin{aligned}
\partial_t A_T(t)=&-(\mu+\gamma_P) A_T(t) + a \frac{S(t)}{S(t)+P(t)+A(t)+A_T(t)}(1-\delta) \left\langle \epsilon(s) s \left[1-\left(1-\lambda \frac{P(t)}{S(t)+P(t)+A(t)+A_T(t)}\right)^{s-1}\right]\right\rangle\\ 
+& a \frac{A(t)}{S(t)+P(t)+A(t)+A_T(t)} \left\langle \epsilon(s) s \left[1-\left(1-\lambda \delta \frac{S(t)}{S(t)+P(t)+A(t)+A_T(t)}\right)^{s-1}\right]\right\rangle \\
+& a \frac{S(t)}{S(t)+Y(t)} (1-\delta) \left\langle \epsilon(s) s \left[1-\left(1-\lambda \frac{A(t)+A_T(t)}{S(t)+Y(t)}\right)^{s-1}\right]\left[1-\left(1-\lambda \delta \frac{S(t)}{S(t)+Y(t)}\right)^{s-2}\right]\right\rangle
\end{aligned}
\end{equation}
where the first term on the right hand side accounts for spontaneous recovery and for isolation of traced asymptomatic, the second, third and fourth terms account respectively for forward, backward and sideward CT.

Finally, the equation for the probability $A_Q(t)$ to be in the asymptomatic quarantined state is
\begin{equation}
\partial_t A_Q(t)=-\mu A_Q(t) + \gamma_P A_T(t)
\end{equation}
where the first term on right hand side accounts for spontaneous recovery and the second term for isolation of traced asymptomatic nodes. \\

Overall, we obtain a set of $5$ coupled differential non-linear equations:
\begin{align}
\partial_t P(t)&= -\gamma_P P(t) + a \frac{S(t)}{S(t)+Y(t)} \delta \left\langle s \left[1-\left(1-\lambda \frac{Y(t)}{S(t)+Y(t)}\right)^{s-1}\right] \right\rangle\\
\partial_t I(t)&=-\mu_I I(t) + \gamma_P P(t)\\
\partial_t A(t)&=\begin{aligned}[t] -\mu A(t) &+ a \frac{S(t)}{S(t)+Y(t)} (1-\delta) \left\langle s \left[1-\left(1-\lambda \frac{Y(t)}{S(t)+Y(t)}\right)^{s-1}\right]\right\rangle\\
&-C^{Forward}-C^{Backward}-C^{Sideward}\end{aligned}\\
\partial_t A_T(t)&=-(\mu+\gamma_P) A_T(t) +C^{Forward}+C^{Backward}+C^{Sideward} \\
\partial_t A_Q(t)&=-\mu A_Q(t) + \gamma_P A_T(t)
\end{align}
with 
\begin{align}
C^{Forward}&= a \frac{S(t)}{S(t)+Y(t)}(1-\delta) \left\langle \epsilon(s) s \left[1-\left(1-\lambda \frac{P(t)}{S(t)+Y(t)}\right)^{s-1}\right]\right\rangle \\
C^{Backward}&= a \frac{A(t)}{S(t)+Y(t)} \left\langle \epsilon(s) s \left[1-\left(1-\lambda \delta \frac{S(t)}{S(t)+Y(t)}\right)^{s-1}\right]\right\rangle \\
C^{Sideward}&= a \frac{S(t)}{S(t)+Y(t)} (1-\delta) \left\langle \epsilon(s) s \left[1-\left(1-\lambda \frac{A(t)+A_T(t)}{S(t)+Y(t)}\right)^{s-1}\right] \left[1-\left(1-\lambda \delta \frac{S(t)}{S(t)+Y(t)}\right)^{s-2}\right]\right\rangle
\end{align}
where $S(t)=1-P(t)-I(t)-A(t)-A_T(t)-A_Q(t)$ and $Y(t)=P(t)+A(t)+A_T(t)$.

This set of equations admits as a stationary state the absorbing state, a configuration where all the population is susceptible. To obtain the condition for the stability of the absorbing state, i.e. the epidemic threshold, we apply a linear stability analysis around the absorbing state.
We consider $\lambda$ as the control parameter and the epidemic threshold is $\lambda_C$.

We neglect the second order terms in probabilities and we obtain a linearized set of $5$ differential equations:

\begin{align}
\partial_t P(t)&= -\gamma_P P(t) + \lambda \delta a \langle s(s-1) \rangle [P(t)+A(t)+A_T(t)]\\
\partial_t I(t)&=-\mu_I I(t) + \gamma_P P(t)\\
\partial_t A(t)&=-\mu A(t)\begin{aligned}[t]&+\lambda (1-\delta) a \langle s(s-1) \rangle [P(t)+A(t)+A_T(t)] \\
&-C^{Forward}-C^{Backward}-C^{Sideward}\end{aligned}\\
\partial_t A_T(t)&=-(\mu+\gamma_P) A_T(t) +C^{Forward}+C^{Backward}+C^{Sideward}\\
\partial_t A_Q(t)&=-\mu A_Q(t) + \gamma_P A_T(t)
\end{align}
where the linearized CT terms:
\begin{align}
C^{Forward}&=\lambda (1-\delta) a \langle \epsilon(s) s (s-1) \rangle P(t) \\
C^{Backward}&= a \left\langle \epsilon(s) s \left[1-(1-\lambda \delta)^{s-1}\right] \right\rangle A(t) \\
C^{Sideward}&= \lambda (1-\delta) a \left\langle \epsilon(s) s(s-1) \left[1-(1-\lambda \delta)^{s-2}\right]\right\rangle \left[A(t)+A_T(t)\right]
\end{align}

The quantity $\overline{n} = a \langle s(s-1) \rangle$ is the average
number of links established by an individual per unit of time.
We will consider this quantity as a constant. Making the dependence on
$\overline{n}$ explicit, the linearized equations read:

\begin{align}
\partial_t P(t)&= -\gamma_P P(t) + \lambda \delta \overline{n} [P(t)+A(t)+A_T(t)]\\
\partial_t I(t)&=-\mu_I I(t) + \gamma_P P(t)\\
\partial_t A(t)&=-\mu A(t)\begin{aligned}[t]&+\lambda (1-\delta) \overline{n} [P(t)+A(t)+A_T(t)] \\
&-C^{Forward}-C^{Backward}-C^{Sideward}\end{aligned}\\
\partial_t A_T(t)&=-(\mu+\gamma_P) A_T(t) +C^{Forward}+C^{Backward}+C^{Sideward}\\
\partial_t A_Q(t)&=-\mu A_Q(t) + \gamma_P A_T(t)
\end{align}

where the linearized CT terms are:
\begin{align}
C^{Forward}&=\lambda (1-\delta) \overline{n} \frac{\langle \epsilon(s) s (s-1) \rangle}{\langle s(s-1) \rangle} P(t) \\
C^{Backward}&= \frac{\overline{n}}{\langle s(s-1) \rangle} \left\langle \epsilon(s) s \left[1-(1-\lambda \delta)^{s-1}\right] \right\rangle A(t)\\
C^{Sideward}&=\lambda (1-\delta) \frac{\overline{n}}{\langle s(s-1) \rangle} \left\langle \epsilon(s) s(s-1) \left[1-(1-\lambda \delta)^{s-2}\right]\right\rangle \left[A(t)+A_T(t)\right]
\end{align}

The set of linearized equations can be written as:
\begin{equation}
\begin{bmatrix}
\partial_t A_Q(t)\\
\partial_t I(t)\\
\partial_t P(t)\\
\partial_t A(t)\\
\partial_t A_T(t)
\end{bmatrix}
=J \begin{bmatrix}
A_Q(t)\\
I(t)\\
P(t)\\
A(t)\\
A_T(t)
\end{bmatrix}
\end{equation}

where $J$, the Jacobian matrix of this set of $5$ linearized equations, is:

\begin{equation}
J=\begin{bmatrix}
-\mu & 0 & 0 & 0 & \gamma_P\\
0 & -\mu_I & \gamma_P & 0 & 0\\
0 & 0 & -\gamma_P+\beta & \beta & \beta\\
0 & 0 & \Delta \left(1-\frac{\langle \epsilon(s) s(s-1) \rangle}{\langle s(s-1) \rangle} \right) & -\mu+\Delta-\Gamma-\Phi & \Delta-\Phi\\
0 & 0 & \Delta \frac{\langle \epsilon(s) s(s-1) \rangle}{\langle s(s-1) \rangle} & +\Gamma+\Phi & -\mu-\gamma_P+\Phi\\
\end{bmatrix}
=
\begin{bmatrix}
\mathbb{A}(2\text{x}2) & \mathbb{C}(2\text{x}3) \\
\mathbb{O}(3\text{x}2) & \mathbb{B}(3\text{x}3)
\end{bmatrix}
\label{eq:J}
\end{equation}

where $\Phi=\lambda (1-\delta) \frac{\overline{n}}{\langle s(s-1) \rangle} \left\langle \epsilon(s) s(s-1) \left[1-(1-\lambda \delta)^{s-2}\right] \right\rangle$, $\Gamma= \frac{\overline{n}}{\langle s(s-1) \rangle} \left\langle \epsilon(s) s \left[1-(1-\lambda \delta)^{s-1}\right] \right\rangle$, $\beta=\lambda \delta \overline{n}$ and $\Delta=\lambda (1-\delta) \overline{n}$. \\

The condition for the
stability of the absorbing state is obtained by imposing all
eigenvalues of $J$ to be negative.
The Jacobian matrix is a block matrix and hence
we can consider separately the two blocks
on the diagonal: for the first block $\mathbb{A}$ it is evident that
the eigenvalues, $\xi_{1}=-\mu$, $\xi_2=-\mu_I$, are all negative.
Therefore, it is sufficient to study block $\mathbb{B}$,
which is a $3\text{x}3$ matrix. The epidemic threshold is therefore
obtained by numerically diagonalizing the matrix $\mathbb{B}$ and
imposing all its eigenvalues to be negative.

\subsection{Limit cases}
The mean-field equations and the epidemic threshold are obtained for arbitrary $\Psi(s)$ and for completely general $\epsilon(s)$: this allows to introduce complicated effects, such as heterogeneity in simplices size and CT strategies on classes of simplices. 

Due to the complicated structure of the conditions for the stability, it is possible to derive the epidemic threshold $\lambda_C$ only by solving the stability conditions numerically. However, there are some limit cases in which the equations are considerably simplified, allowing to obtain the epidemic threshold in an explicit analytic form.

Note that the epidemic threshold $\lambda_C$ is still finite even if $\epsilon(s)=1$ $\forall s$, due to the presymptomatic infections and to the delay in isolation of traced nodes. 

\subsubsection{Non-adaptive case (NA)}
Here we consider the non-adaptive case, in which no adaptive behaviour is implemented, i.e. infected nodes behave as if they were susceptible $b_I=b_S=b$. Thus, in this case $\epsilon(s)=0$, $\forall s$ and $\gamma_P/\mu=1$.
Considering these values we obtain the following equation for the critical condition:

$$-\mu+\lambda a \langle s(s-1) \rangle=0$$

So we obtain an explicit form for the epidemic threshold $\lambda_C$ in the non-adaptive case:

\begin{equation}
\lambda_C^{NA}=\frac{\mu}{a \langle s(s-1) \rangle}=\frac{\mu}{\overline{n}}
\label{eq:Na}
\end{equation}
which is Eq.~(1) in the main paper. It reproduces the results previously obtained in Ref.~\cite{petri2018simplicial} and for only pairwise interactions $\Psi(s) = \delta(s-2)$ it reproduces the results previously obtained in Refs.~\cite{Pozzana2017,Mancastroppa2020,mancastroppa2021contacttracing}.

\subsubsection{Isolation of only symptomatic nodes}
Here we consider the case in which only symptomatic nodes are isolated
as soon as they develop symptoms, i.e. no CT is implemented. Thus, in
this case $\epsilon(s)=0$, $\forall s$, while $b_I=0$. Considering
these values we obtain the following equation for the critical
condition:

$$-\gamma_P \mu +\lambda \overline{n} (\delta \mu + (1-\delta) \gamma_P)=0$$

So we obtain an explicit form for the epidemic threshold $\lambda_C$:

\begin{equation}
\lambda_C^{sympto}=\lambda_C^{NA} \frac{\frac{\gamma_P}{\mu}}{\delta+(1-\delta)\frac{\gamma_P}{\mu}}
\end{equation}
which is Eq.~(2) in the main paper. For pairwise interactions
$\Psi(s)=\delta(s-2)$ it reproduces the results previously obtained in
Ref.~\cite{mancastroppa2021contacttracing}, and for $\gamma_P/\mu=1$
it reproduces the NA case (Eq. \eqref{eq:Na}).

\subsubsection{Homogeneous case}
Here we consider the case in which the size of simplices is homogeneous, i.e. $\Psi(s) = \delta(s-\overline{s})$, with constant probability for a simplex to be traced with CT, i.e. $\epsilon(s)=\epsilon$. 

If we consider $\overline{s}=2$, that is only pairwise interactions, we obtain a quadratic equation in $\lambda$ for the critical condition:

$$\frac{\lambda^2}{\mu^2} \overline{n}^2 \delta^2 \epsilon + \frac{\lambda}{\mu} \overline{n} \left(\delta+(1-\delta-\delta \epsilon) \frac{\gamma_P}{\mu}\right) - \frac{\gamma_P}{\mu}=0$$

The equation can be solved and we obtain:

\begin{equation}
\lambda_C^{\overline{s}=2}=\lambda_C^{NA} \frac{2 \frac{\gamma_P}{\mu}}{\delta+(1-\delta-\epsilon \delta)\frac{\gamma_P}{\mu}+\sqrt{(\delta+(1-\delta-\epsilon \delta)\frac{\gamma_P}{\mu})^2+4 \delta^2 \epsilon \frac{\gamma_P}{\mu}}}
\label{eq:s2}
\end{equation}
which reproduces the results obtained in Ref.~\cite{mancastroppa2021contacttracing}.

If we consider $\overline{s} \to \infty$, which means to consider only the activation of simplices in which all nodes participate, we obtain a linear equation in $\lambda$:

$$\lambda [ \delta \mu \overline{n}(\gamma_P+\mu) + \overline{n} \gamma_P (1-\delta) (\mu+\gamma_P(1-\epsilon))]-\gamma_P \mu (\mu+\gamma_P)=0$$

The equation can be solved and we obtain:

\begin{equation}
\lambda_C^{\overline{s} \to \infty}=\lambda_C^{NA} \frac{\frac{\gamma_P}{\mu}(\gamma_P +\mu)}{\delta(\gamma_P+\mu)+ \gamma_P (1-\delta) (1+\frac{\gamma_P}{\mu}(1-\epsilon))}
\label{eq:sinf}
\end{equation}

The maximum epidemic threshold $\lambda_C^{max}$ is obtained for $\overline{s} \to \infty$ and $\epsilon=1$: indeed, in that case all nodes are traced at their infection, both if infected by $A$ or $A_T$, through sideward CT, and if infected by $P$, through forward CT. In this case the epidemic threshold is still finite since isolation of traced individuals occurs with a delay $\tau_P$ and because of the presymptomatic infection:

\begin{equation}
\lambda_C^{max}= \lambda_C^{NA} \frac{\frac{\gamma_P}{\mu}(\gamma_P+\mu)}{\gamma_P+\delta \mu}
\label{eq:l_max}
\end{equation}

\section{Supplementary Method 2: WIFI data for the University of Parma} 

Passive WIFI data require preprocessing before any data analysis and, in particular, to obtain the empirical distributions $\Psi(s)$ used in the main text. 
The staff of the "ICT  services" office (ICTS) of Parma University extracts a tabular file (called log file) from the login management system every day containing all daily connections to the WIFI 
networks installed in all university buildings.
The login management system writes a line to the log file whenever a network connection begins or ends.
 Each line contains user information, device
information and session information. To reconstruct all user connections from the log file, the following attributes are used: 
\begin{itemize}
 \item \textbf{Username}: user's email address;
 \item \textbf{Type of user}: this attribute allows us to distinguish the users in students, structured staff and external guests;
 \item \textbf{Calling device ID}: MAC address of the device to distinguish all user's devices;
 \item \textbf{Device type}: this attribute distinguishes the type of user's devices (computers, smartphones or tablets);
 \item \textbf{Called station ID}: MAC address of the access point (AP) to which the device asks to connect;
 \item \textbf{Status type}: this attribute indicates whether this accounting request marks the beginning of the user service (\emph{Start}) or the end (\emph{Stop});
 \item \textbf{Date-time}: this attribute represents the day and the time for this accounting request;
 \item \textbf{Session ID}: this attribute is a unique accounting ID to make it easy to match start and stop record in a log file.
\end{itemize}

{\bf Privacy-preservation Mechanisms}. Due to the European regulation on privacy (GDPR) we cannot use directly the log files as they contain personal data. We conducted a Data Protection Impact Assessment in order to fulfill data minimization principles and be compliant with the latest Regulation on Privacy and Electronic Communications.
Accordingly, we developed a procedure (script) to be run within the ICTS domain that completely anonymizes the data and compute the aggregated quantities described in the paper:

\begin{enumerate}
\item the script removes all the personal data, such as name and device information, and implements a script which replaces the personal data with random 16-digit hexadecimal strings (pseudo-anonymizing the data). In order for the data to be consistent after being pseudo-anonymized, all correlations between attributes for all lines in log file are maintained. Every time a new personal data is replaced with a random string, ICTS saves the correspondence \emph{personal data-string} into \emph{keys files} and uses it every time that sensitive data is found. The seed for random string generation is changed every 24-hours.  
\item in the dataset some connections are interrupted for short times due to various reason (such as weak signal or user's device in standby). Hence, if two or more consecutive connections of a single user to the same APs are present and the interval time  $\tau$  between them is less then 5 minutes,  a single connection is considered with start at the first connection and stop at the last one.
\item the script computes the number of devices $s$ connected to a given AP for each 15 minutes. Devices must remain connected to the AP for the whole time. We call $s$ the cluster size at the corresponding AP at a given time interval.
\item the ICTS script provides two measures of aggregated quantities, which are the only data we can access directly:
\begin{itemize}
\item {\it Presences}: the total number of individual connected to the University WIFI during each 15 minutes interval.     \item {\it Group Size Statistic}: the total number of 15 minutes clusters of a given size $s$ present in the whole University
during each day of observation.    
\end{itemize}
\item All temporary data are deleted after 24 hours and only the above measures are retained. 
\end{enumerate}

The data of the presences in the University Campus has been plotted in Fig. 5\textbf{b} of the main text. The daily group size distribution has been used to obtain the distributions $\Psi(s)$ in Fig. 4\textbf{a} of the main text. 

\section{Supplementary Note 1: Effects of the fraction of symptomatic individuals}

\begin{figure*}
\centering
\includegraphics[width=180mm]{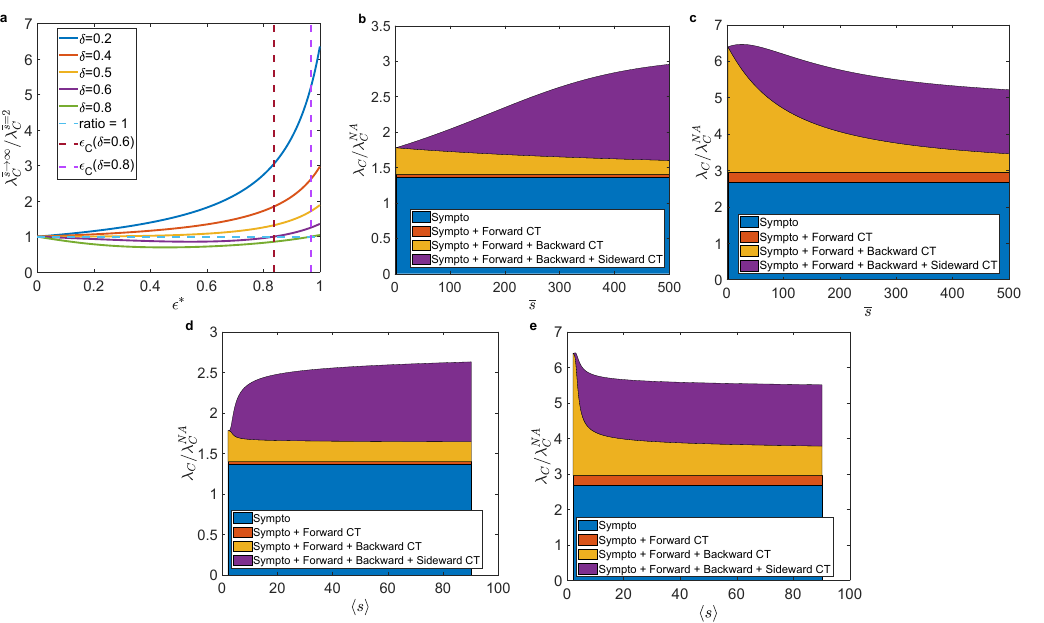}
 \caption{{\bf Effects of the fraction of symptomatic infections.} \textbf{a} By considering $\Psi(s)=\delta(s-\overline{s})$, we plot the ratio between the epidemic thresholds $\lambda_C^{\overline{s} \to \infty}$ and $\lambda_C^{\overline{s}=2}$ as a function of $\epsilon^*=\epsilon$, for several fixed $\delta$ values. The vertical dashed lines indicate the $\epsilon_C$ value for each $\delta>1/2$ value considered. \textbf{b} For the same $\Psi(s)$ of panel \textbf{a}, we plot the ratio $\lambda_C/\lambda_C^{NA}$ as a function of $\overline{s}$ for symptomatic isolation and activating progressively all CT mechanisms, showing their relative contribution to the epidemic threshold when all are active. In this case $\delta=0.3$ and $\epsilon=0.6$. \textbf{c} Same of panel \textbf{b} with $\delta=0.7$. \textbf{d}-\textbf{e} Same of panels \textbf{b}-\textbf{c} with $\Psi(s) \sim s^{-(\nu+1)}$, $s \in [2, 500]$ and $\epsilon(s)=\epsilon^*=0.6 \, \forall s$. In all panels the other parameters are fixed as discussed in the Methods of the main text.}
 \label{fig:figure2}
\end{figure*}

Our model allows to study how the CT mechanisms depend on the parameters of the epidemics. In particular we show that the fraction of symptomatic infections $\delta$, i.e. the fraction of individuals who develop symptoms, has a relevant effect on the relative contribution of the CT mechanisms. 
First, the presence of a high fraction of symptomatic individuals induces an increase in the contribution of forward CT, since it traces infections produced by presymptomatic individuals. Moreover, the value of $\delta$ has an impact on the contribution of backward and sideward CT. This is evident by considering the homogeneous case $\Psi(s) = \delta(s-\overline{s})$ and comparing the $\overline{s}=2$ case, where sideward tracing is absent, with the $\overline{s} \to \infty$ system, where sideward tracing dominates.
By considering Eqs. \eqref{eq:s2}-\eqref{eq:sinf}, it can be shown that for $\delta<1/2$ we have:
\begin{equation}
\lambda_C^{\overline{s}=2}<\lambda_C^{\overline{s} \to \infty}
\end{equation}
Indeed for infectious diseases with a high rate of asymptomatics~\cite{Fraser2004} (such as COVID-19 and its variants~\cite{Lavezzo2020}) sideward tracing becomes considerably more effective than backward tracing and CT typically occurs on large simplices. 

Instead, by considering Eqs. \eqref{eq:s2}-\eqref{eq:sinf} for $\delta>1/2$ we have:
\begin{equation}
\begin{cases}
\lambda_C^{\overline{s}=2}\geq\lambda_C^{\overline{s} \to \infty} \,\,\, \text{if} \,\,\, \epsilon\leq\epsilon_C\\
\lambda_C^{\overline{s}=2}<\lambda_C^{\overline{s} \to \infty} \,\,\, \text{if} \,\,\, \epsilon>\epsilon_C\\
\end{cases}
\end{equation}
with 
\begin{equation}
\epsilon_C=\frac{(\gamma_P+\mu)(1-2\delta)}{(1-2\delta)\gamma_P-\delta \mu}
\end{equation}
In this case, for small $\epsilon<\epsilon_C$ the backward CT dominates due to the high fraction of symptomatics, while if $\epsilon>\epsilon_C$, sideward tracing is more effective. 
The behavior of $\lambda_C^{\overline{s} \to \infty}$ and $\lambda_C^{\overline{s}=2}$ as a function of $\epsilon$ and $\delta$ is summarized in Supplementary Fig.~\ref{fig:figure2}{\bf a}. Note that for $\delta$ larger than $1/2$ sideward tracing on large simplices becomes dominant only for extremely large values of $\epsilon$, since $\epsilon_C$ increases rapidly with $\delta$ and $\epsilon_C \sim 1$. 

In Supplementary Fig.~\ref{fig:figure2}{\bf b-c}, again in the case with a single cluster size $\overline{s}$, we plot the gain in the critical infection probability $\lambda_C/\lambda_C^{NA}$ as a function of $\overline{s}$ and we show the contribution of the different tracing mechanisms.
For small $\delta<1/2$ (many asymptomatic infections) tracing is more effective at large cluster sizes and the most relevant mechanism is the sideward CT, while for large 
$\delta>1/2$ (many symptomatic infections) the peak in $\lambda_C/\lambda_C^{NA}$ is obtained at small $\overline{s}$ where the most relevant contribution is given by backward tracing. An analogous result is obtained in Supplementary Fig.~\ref{fig:figure2}{\bf d-e} where we consider a power law distribution $\Psi(s)$ of cluster sizes with different average size $\langle s \rangle$.

\section{Supplementary Note 2: Stability of the results}

Let us now show that our results are stable when we change some relevant parameters in the modelling scheme. 

\subsection{Role of the recovery rate}

\begin{figure}[t!]
\centering
\includegraphics[width=180mm]{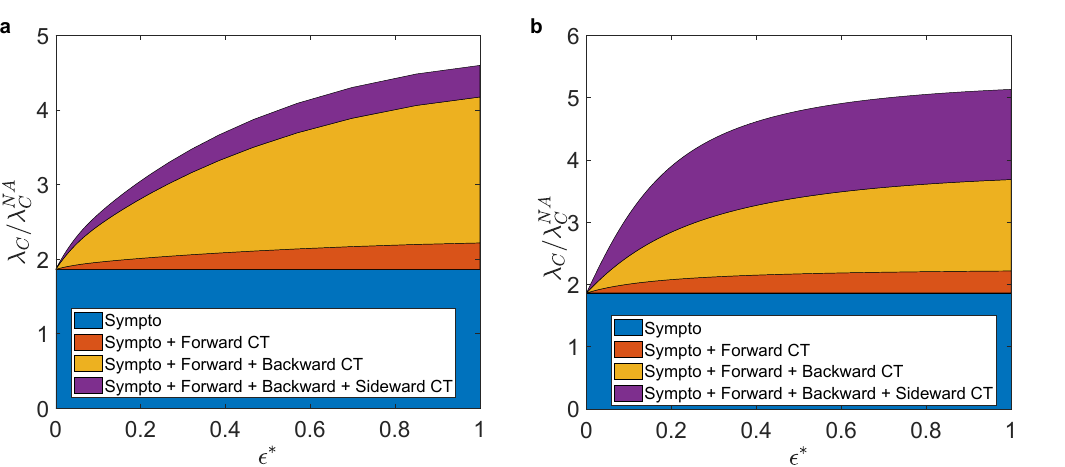}
 \caption{{\bf Role of the recovery rate.} \textbf{a} We plot the ratio $\lambda_C/\lambda_C^{NA}$ as a function of $\epsilon^*$ for symptomatic isolation and activating progressively all CT mechanisms, showing their relative contribution to the epidemic threshold when they all are active, considering the empirical $\Psi(s)$ during the partial opening phase and implementing a tracing strategy targeted at large simplices $\epsilon(s)=\theta(s-s^*)$. \textbf{b} Same as panel \textbf{a} with $\Psi(s) \sim s^{-(\nu+1)}$, $s \in [2,200]$ and $\nu=1.5$. In all panels $\tau=8$ days and the other parameters are fixed as discussed in the Methods of the main text.}
 \label{fig:figure3}
\end{figure}

A comparison of Supplementary Fig.~\ref{fig:figure3} with Fig. 4 of the main text shows that the contribution of the different CT mechanisms does not change qualitatively when varying the recovery rate $\mu=1/\tau$. Note that this is not a trivial property, since in our model $\mu$ is not a simple scaling factor of the critical infection probability $\lambda_C$ as it occurs typically in epidemics on dynamical networks~\cite{Mancastroppa2019,Mancastroppa2020,mancastroppa2021contacttracing}. This is due to the fact that $C^{Backward}$ and $C^{Sideward}$ are non linear functions of $\lambda$ and therefore the sign of the eigenvalues of the stability Jacobian matrix (Eq. \eqref{eq:J}) does not depend simply on the ratio $\lambda/\mu$. Clearly a linear expression for $C^{Backward}$ and $C^{Sideward}$ is recovered for $\Psi(s) = \delta(s-2)$.

\subsection{Effects of errors in the targeted tracing strategy}

\begin{figure}[t!]
\centering
\includegraphics[width=180mm]{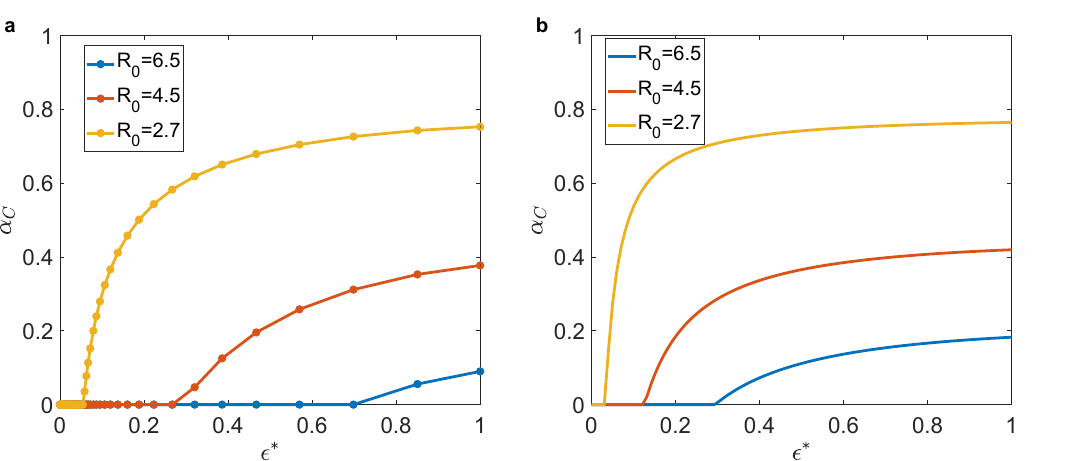}
 \caption{{\bf Effects of errors in the tracing strategy targeted on large simplices.} \textbf{a} We consider the targeted CT strategy on large simplices with error $\epsilon(s)=(1-\alpha) \theta(s-s^*)$: we plot the critical value $\alpha_C$, i.e. the critical $\alpha$ which bring the system above the epidemic threshold, as a function of $\epsilon^*$ that is the average fraction of nodes traced per simplex when $\alpha=0$ and it fixes $s^*$. The curve is plotted for several $R_0$ values and considering the empirical $\Psi(s)$ during the partial opening phase. \textbf{b} Same of panel \textbf{a} but considering a synthetic distribution $\Psi(s) \sim s^{-(\nu+1)}$ with $\nu=1.5$, $s \in [2,200]$. In all panels the other parameters are fixed as discussed in the Methods of the main text.}
 \label{fig:figure4}
\end{figure}

The targeted tracing strategy requires to perfectly trace large simplices, however this is a hard task in real situations and typically a fraction $\alpha$ of the simplices cannot be traced. We investigate the effect of such a small untraced fraction, by considering $\epsilon(s)=(1-\alpha) \theta(s-s^*)$ and keeping fixed $s^*$, so that only simplices with $s \geq s^*$ are traced with probability $(1-\alpha)$. In this case $\alpha \in [0,1]$: for $\alpha=0$ the targeted tracing is perfect and the average fraction of nodes traced per simplex is $\frac{\langle \epsilon(s) (s-1)\rangle}{\langle s-1 \rangle}=\epsilon^*$; increasing $\alpha$ we introduce an error in the tracing of large simplices and the fraction of traced nodes is reduced $\frac{\langle \epsilon(s) (s-1)\rangle}{\langle s-1 \rangle}<\epsilon^*$; for $\alpha=1$ no tracing is performed and $\epsilon(s)=0$ $\forall s$. 
The results are summarized in Supplementary Fig.~\ref{fig:figure4} where we fix $R_0=\lambda/\lambda_C^{NA}$ at values for which the adaptive system without tracing is in the active phase. Then we plot as a function of $\epsilon^*$ the critical value $\alpha=\alpha_C$ which is able to keep the system in the active phase. Clearly for small enough $\epsilon^*$ the system remains active even for $\alpha=0$ (in that case we indicate $\alpha_C=0$). For larger $\epsilon^*$ the system is in the absorbing phase at $\alpha=0$ and a finite value of $\alpha=\alpha_C>0$ is necessary to generate an active system.
We perform this analysis both for empirical and synthetic distribution $\Psi(s)$. 
Supplementary Fig.~\ref{fig:figure4} points out that $\alpha_C$ increases rapidly with $\epsilon^*$ (slope of the curve), indicating a significant stability of our results on the effectiveness of the targeted tracing strategy: indeed if the system is well inside the absorbing phase (sufficiently high $\epsilon^*$) then for reasonable values of $\alpha$ (e.g $\alpha=0.1$) the system remains in the absorbing phase and the epidemic does not spread.
